\documentclass[12pt, draftclsnofoot, onecolumn]{IEEEtran}
\usepackage{bm,cite,float,amsmath,amssymb,amsthm}

\usepackage{algorithm}
\usepackage[noend]{algpseudocode}
\usepackage{indentfirst}
\usepackage{xcolor}
\usepackage{dsfont}
\usepackage{nicefrac} 

\makeatletter
\def\BState{\State\hskip-\ALG@thistlm}
\makeatother

\usepackage{amssymb}
\usepackage{amsmath}
\usepackage{graphicx}
\usepackage{cite}
\usepackage{citesort}
\usepackage{balance}
\usepackage[utf8]{inputenc}
\usepackage{makecell}
\usepackage{caption}

\IEEEoverridecommandlockouts

\usepackage{amsthm}
\usepackage{graphicx,epstopdf}
\usepackage{epsfig}	
\usepackage{amsfonts,balance}
\usepackage{bbm}
\floatname{algorithm}{Algorithm}
\usepackage{lipsum}

\newtheorem{theorem}{Theorem}

\newtheorem{corollary}{Corollary}

\makeatletter
\def\ScaleIfNeeded{%
\ifdim\Gin@nat@width>\linewidth \linewidth \else \Gin@nat@width
\fi } \makeatother

\begin{document}

\title{Weighted Sum-Rate Maximization for Rate-Splitting Multiple Access Based Multi-antenna Broadcast Channel with Confidential Messages}

\author{Huiyun~Xia, ~\IEEEmembership{Student~Member,~IEEE,}
		Yijie~Mao, ~\IEEEmembership{Member,~IEEE,}
        Xiaokang~Zhou, ~\IEEEmembership{Student Member,~IEEE,}
		Bruno~Clerckx, ~\IEEEmembership{Fellow,~IEEE,}        
        Shuai~Han, ~\IEEEmembership{Senior Member,~IEEE,}
        and Cheng Li,~\IEEEmembership{Senior Member,~IEEE}.

\vspace{-0.2cm}
\thanks{This work was supported by the Natural Science Foundation of China under Grant 61831002. The work of Huiyun Xia was supported by the China Scholarship Council. A preliminary version of this work has been accepted by the 2022 IEEE Wireless Communications and Networking Conferences Workshop \cite{Xia2022secrecyRS}. (Corresponding author: Shuai Han)}
\thanks{
Huiyun~Xia, Xiaokang~Zhou and Shuai~Han are with the School of Electronics and Information Engineering, Harbin Institute of Technology, Harbin, 150001, P. R. China (emails: summerxiahy @163.com; kangsenneo@sina.com; hanshuai@hit.edu.cn). }
\thanks{Yijie Mao is with the School of Information Science and Technology, ShanghaiTech University, Shanghai, China (email: maoyj@shanghaitech.edu.cn).}
\thanks{Bruno Clerckx is with the Department of Electrical and Electronic Engineering, Imperial College London, London, UK (email: b.clerckx@imperial.ac.uk).}
\thanks{Cheng Li is with the Electrical and Computer Engineering Faculty of Engineering and Applied Science, Memorial University, St. John's, A1B 3X5, Canada (email: licheng@mun.ca).}
}

\maketitle

\begin{abstract}
In this paper, we take all users' secrecy rate requirements into consideration and propose a rate-splitting multiple access (RSMA)-based secure beamforming approach to maximize the weighted sum-rate (WSR), where multi-user interference (MUI) is partially decoded and partially treated as noise.
User messages are split and encoded into common and private streams.
Each user not only decodes the common stream and the intended private stream, but also eavesdrops other users' private streams.
To solve the introduced non-convex security constraints in the formulated problem, a successive convex approximation (SCA)-based approach is first proposed to maximize the instantaneous WSR under perfect channel state information at the transmitter (CSIT). 
When only imperfect CSIT is available, to ensure robustness, a weighted ergodic sum-rate maximization problem is formulated. Then, a joint weighted minimum mean square error and SCA-based alternating optimization algorithm is proposed to optimize the precoder.
Numerical results demonstrate that, compared with baseline schemes, the performance advantage of the proposed RSMA-based secure beamforming design has been shown in terms of WSR and robustness to channel errors while ensuring all users' security requirements thanks to its powerful interference management capability. 
\end{abstract}


\maketitle

\section{Introduction}

\IEEEPARstart{R}{ecently,}  rate-splitting multiple access (RSMA), built upon the rate-splitting (RS) technique, has emerged as a powerful non-orthogonal transmission framework and interference management strategy for wireless networks \cite{Bruno2016RS}. 
RSMA enables user messages to be split into certain common and private parts, after which the private parts are independently encoded into private streams while the common parts are jointly encoded into common streams superposed on top of the private streams.
The superposed streams are then precoded and transmitted from the multi-antenna transmitter. 
At the receiver side, successive interference cancellation (SIC) is applied at each user to enable sequential decoding of the common streams and the intended private stream. 
Receivers then reconstruct the original messages by extracting the intended common part from the decoded common message and combine it with the decoded private message \cite{Bruno2016RS}. 
RSMA enables a more flexible interference management approach of partially decoding the interference and partially treating the interference as noise, and it has been shown to bridge and outperform space division multiple access (SDMA) and power-domain non-orthogonal multiple access (NOMA) \cite{mao2018rsma}.
Furthermore, the performance advantage of RSMA has also been demonstrated in terms of robustness against imperfect CSIT \cite{Hamdi2016robustRS} and user mobility \cite{onur2021mobility}, SE \cite{mao2018rsma}, EE \cite{mao2019RSSEEE}, user fairness \cite{Chen2020userfairness}, reliability \cite{Caus2018reliability}, QoS enhancements \cite{mao2018rsma}, etc. 

Due to the broadcast nature, wireless communications is naturally vulnerable to security breaches. 
Since RSMA is capable of serving multiple users using the same time-frequency resource, it is highly possible that when the users in RSMA communication systems receive their intended messages, they may also wiretap messages intended for other co-channel users. This situation occurs when the potential eavesdropper is a registered user served by the transmitter but has a lower authentication level or malicious intention to wiretap messages delivered to other users.
Such threat of information leakage is especially fatal in RSMA communications, since the common stream in which is prone to eavesdrop attack by other undesired users. Therefore, it is necessary to reap the benefits of RSMA in SE while maintaining its secrecy. 

Physical layer security (PLS) takes advantage of the intrinsic randomness and fading characteristics of wireless channels and has exhibited substantial performance merits over traditional cryptographic encryption methods in terms of computational complexity and adaptability to heterogeneous environments. Other than conventional cryptographic solutions, which require computation-demanding encryption algorithms and intricate key management and distribution strategies \cite{xie2021PLAsurvey}, physical layer security adopts advanced signal processing techniques to guarantee perfect communication security from an information-theoretic perspective and has become a hot spot field of research for complementing data encryption in the application layer \cite{Yu2020IRSsecurity}.  
The fundamental principle of physical layer security is to widen the performance gap between legitimate users and illegal users by exploiting the intrinsic randomness of transmission medium in data transmission.
First proposed by Shannon \cite{shannon1949secrecy} and later developed by Wyner \cite{wyner1975wiretap}, physical layer security has evolved from point-to-point systems \cite{Csiszar1978BC,Parada2005P2P} to multi-user systems \cite{Geraci2012RCI}, where multi-antenna techniques are adopted to leverage the advantage in spatial multiplexing and diversity, and meanwhile to provide sufficient spatial freedom for separating different signals in the same time-frequency resource.  
Recently, the investigation on RSMA-assisted physical layer security mainly focus on taking advantage of the dual functions of the common stream, using the common stream serving as both useful data for legitimate users and jamming signals for adversarial users. 
For example, Hao et al \cite{hao2020robustsecureRS} studied RSMA-based two-user multiple input single output (MISO) secure communications and designed a robust and secure resource allocation strategy for user fairness optimization. 
Ping et al \cite{Ping2020cooperativeRS} investigated the cooperative rate-splitting (CRS) technique, which is originally proposed in \cite{Zhang2019CRS}, and proposed a CRS strategy to combat the external single-antenna eavesdropper in a MISO broadcast channel (BC), where the proposed strategy considered the legitimate user to opportunistically relay and forward its re-encoded common message to serve as AN to confuse the eavesdropper and as a useful message for the other legitimate user. 
The precoders and time-slot allocation strategy were jointly optimized to maximize the secure sum-rate. 
Lu et al \cite{lu2021secureSWIPT} extended RSMA-based physical layer security to simultaneous wireless information and power transfer (SWIPT) systems and proposed a robust beamforming design to maximize the worst-case EE performance. 
With the aid of the AN\footnote{ Here, the AN does not carry useful information for legitimate users, and it differs from the AN adopted in existing RSMA-based secrecy works \cite{hao2020robustsecureRS,Ping2020cooperativeRS,lu2021secureSWIPT}, where AN not only carries useful information for legitimate users but uses the information to confuse eavesdroppers.}, a RSMA-based secure beamforming and power allocation design was investigated in \cite{cai2021secureRA} to maximize the secure sum-rate.
However, to our best knowledge, 
most current studies focus on anti-eavesdropping scenarios, where the eavesdroppers are not the intended recipients of the messages sent from the transmitter. They solely intercept confidential messages sent to other authorized users.
The performance of RSMA-based secure designs when the eavesdroppers not only receive their intended messages but wiretap the messages intended for other co-channel users remain unknown.

In this work, motivated by the aforementioned performance merits of RSMA in physical layer security and the limitations of existing works, we consider a more general setting where each receiver not only decodes the intended messages but acts as an eavesdropper to tap the information of other users. 
We aim at optimizing the precoders to maximize the weighted sum-rate (WSR) while ensuring the secrecy rate constraints of all users. 
Considering the fact that in RSMA, the common stream is shared by all users whereas only private streams can achieve data confidentiality, to ensure the security demands of each user, how to flexibly schedule the common stream and private streams by optimizing precoders should be carefully designed.  
To our best knowledge, this is the first paper investigating the secure beamforming optimization for the RSMA-assisted MISO BC where each user serves as both a legitimate user to decode its intended message and an eavesdropper to wiretap messages intended for other users. The main contributions of this paper are summarized as follows:

\begin{itemize}
\item[$\bullet$] We construct a RSMA-based MISO BC communication model where the security threat comes from the internal legitimate users, i.e., each user not only decodes its intended message but also decodes messages intended for other users.  
We split each user's message into a common part to be decoded by all users and a private part intended for the corresponding user only. Each user sequentially decodes the intended common and private messages, and then tries to tap the messages of other users. 
We then design secure beamformers to maximize the WSR subject to the secrecy rate constraint of each user. 
\item[$\bullet$] When the perfect CSIT is available, the instantaneous WSR maximization problem is formulated subject to the secrecy rate constraint of each user. 
To handle the introduced non-convex security constraints in the formulated problem, 
 a successive convex approximation (SCA)-based approach is proposed to iteratively optimize the precoder. 
\item[$\bullet$] We further consider a practical case when only the imperfect CSIT is available. 
A weighted ergodic sum-rate (WESR) maximization problem is formulated. 
Due to the intractability of the original problem, a weighted average sum-rate (WASR) maximization problem is reformulated by adopting the sample average approximation (SAA) approach. 
To handle the difference-of-convex structure in the reformulated non-convex problem and reduce computational complexity, a joint weighted minimum mean square error (WMMSE) and SCA-based alternating optimization (AO) algorithm is then proposed to optimize the secure precoders.
\item[$\bullet$] Finally, we compare the proposed algorithm with the multi-user linear precoding (MULP), the zero-forcing (ZF)-based RS extending from the algorithm in \cite{bruno2020ZF}, and NOMA. In particular, NOMA cannot even ensure all users' secrecy constraints, since in NOMA, the entire message of one user is mapped into the common stream, which is entirely decoded by other users. Numerical results demonstrate that, 
compared with the baseline algorithm, the proposed RSMA-based secure beamforming design can achieve a higher WSR performance while satisfying all users’ secrecy rate demands and the performance gap enlarges as the transmit SNR increases. 
Besides, compared with underloaded cases, the WSR performance advantage of the proposed RS design over the baseline algorithm is bigger in overloaded cases.
\end{itemize}

The rest of this paper is organized as follows. Section II introduces the system model. 
The SCA-based beamforming approach for perfect CSIT is specified in Section III, followed by the joint WMMSE and SCA based AO algorithm for imperfect CSIT in Section IV. Numerical results are illustrated in Section V. Finally, Section VI concludes this paper.

\emph{Notation:} Boldfaced lowercase letters and uppercase letters denote column vectors and matrices, respectively; $\bf I$ denotes the identity matrix; ${\mathbb{C}}$ denotes the complex set; $(\cdot)^{\text{T}}$ and $(\cdot)^{\text{H}}$ denote the transpose and conjugate transpose operation, respectively; And $\text{Tr}\{\mathbf{A}\}$ denotes the trace of $\mathbf{A}$. 
$\mathrm{E}_{x}{\{\cdot\}}$ denotes the mathematical expectation with respect to the random variable $x$.
Finally, $\mathcal{CN}(0, \sigma_{n}^2)$ denotes the circularly symmetric complex Gaussian (CSCG) distribution with zero mean and variance $\sigma_{n}^2$.

\section{System Model}

\begin{figure*}[t]
\centering
\centerline{\includegraphics[scale=0.6]{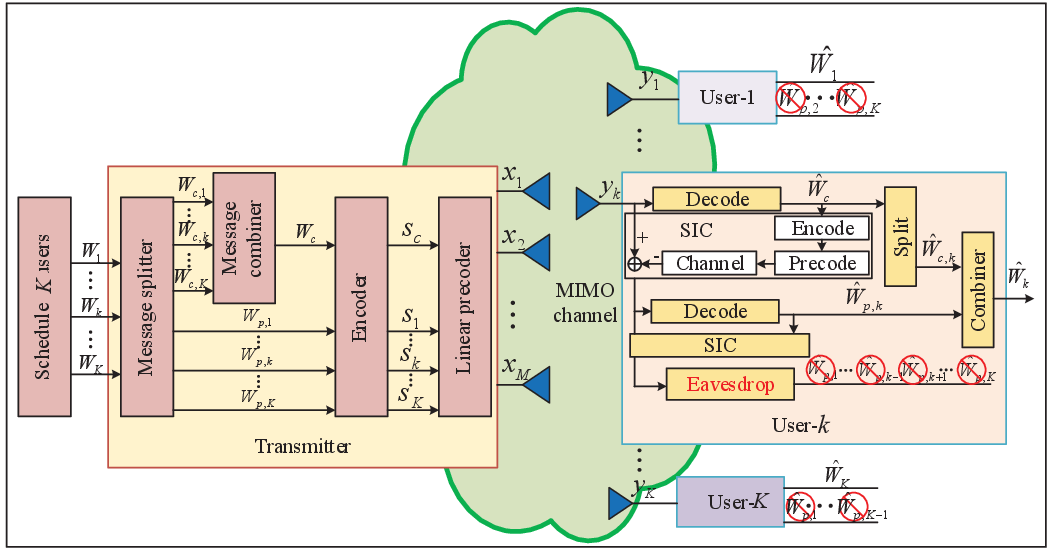}}
\caption{The system model of 1-layer RS based secure MISO BC.}
\label{secure_RSMA_MISO_BC}
\end{figure*}

As illustrated in Fig. \ref{secure_RSMA_MISO_BC}, we consider a downlink RSMA-based MISO BC secure communication model, where one base station (BS) equipped with $N_{\rm t}$ transmit antennas simultaneously serves $K$ single-antenna users, indexed by $\mathcal{K} = \{1,...,K\}$. The BS transmits $K$ confidential messages $W_1,...,W_K$ to the $K$ users sharing the same time-frequency resource. Each message $W_k$ is intended for the $k$th user, denoted by user-$k$, and needs to be kept secret from other users. Each user acts not only as a legitimate user to decode its desired message but as a potential eavesdropper to wiretap the messages transmitted to other users.

Following the principle of 1-layer RS
, the message $W_k$ intended for user-$k$ is split into a common part $W_{\mathrm{c},k}$ and a private part $W_{\mathrm{p},k}$. 
The common parts of all users $W_{\mathrm{c},1}, . . . , W_{\mathrm{c},K}$ are then combined into a common message $W_\mathrm{c}$, which is encoded into the common stream $s_\mathrm{c}$ afterwards using a codebook shared by all users. 
The common stream $s_\mathrm{c}$ is required to be decoded by all users as it contains part of the messages of all users. 
Meanwhile, the private part $W_{\mathrm{p},k}$ of user-$k$ is independently encoded into a private stream $s_k$, which is only required to be decoded by user-$k$. 
The encoded common stream and private streams are then linearly precoded and the resultant transmit signal at BS is
\begin{equation}\label{1}
\mathbf{x}=\mathbf{P}\mathbf{s}=\mathbf{p}_{\mathrm{c}} s_{\mathrm{c}}+\sum\nolimits_{k\in\mathcal{K}}\mathbf{p}_k s_k,
\end{equation}
where ${\bf{s}} = [s_{\mathrm{c}}, s_1, s_2, . . . , s_K]^{\rm{T}}$ is the grouped symbols in a given channel use. 
Assume $\mathrm{E}[\bf{s}\bf{s}^{\rm{H}}] = \bf{I}$. 
With a predefined transmit power limit $P_{max}$, the transmit power constraint is ${\rm{tr}}({\bf P}{\bf P}^{\rm{H}}) \le P_{max}$, where ${\bf P} = [{\bf p}_{\mathrm{c}}, {\bf p}_1, {\bf p}_2, . . . , {\bf p}_K]$ is the integrated precoder matrix and ${\bf p}_{\mathrm{c}}, {\bf p}_k \in {\mathbb C}^{N_t\times1}$ are the precoder for the common stream $s_{\mathrm{c}}$ and the $k$th private stream $s_k$, respectively. 

At user-$k$, the received signal for a given channel state is given by
\begin{align}\label{2}
{y}_k &= \mathbf{h}^\text{H}_k \mathbf{x} + n_k, \quad\forall k \in\mathcal{K},
\end{align}
where $\mathbf{h}_k \in\mathbb{C}^{N_t\times1}$ is the channel between BS and user-$k$. 
We assume the channel state information at the receiver (CSIR) is perfect. Both perfect and imperfect CSIT will be discussed in the following sections.
$n_k\sim\mathcal{CN}(0, \sigma_{n,k}^2)$ is the additive white Gaussian noise (AWGN) at user-$k$. 
Without loss of generality, we assume equal noise variances for all users, i.e., $\sigma_{n,k}^2 = \sigma_{n}^2, \forall k \in \cal{K}$.

At the receiver side, each user first decodes the common stream by treating the interference from all the private streams as noise and then extracts its own part of common message $\widehat W_{{\mathrm{c}},k}$ from $\widehat W_{{\mathrm{c}}}$. 
Assuming Gaussian signalling and perfect SIC, the decoded common message $\widehat W_{{\mathrm{c}},k}$ is then re-encoded, precoded and removed from the received signal. 
After that, each user decodes the intended private stream $s_k$ into $\widehat W_{{\mathrm{p}},k}$ by treating the interference from all other private streams as noise. 
User-$k$ then recovers the original message by combining $\widehat W_{{\mathrm{c}},k}$ with $\widehat W_{{\mathrm{p}},k}$ into $\widehat W_k$.
Consequently, at user-$k$, the corresponding instantaneous achievable rates of common stream $R_{{\mathrm{c}},k}$ and private stream $R_{{\mathrm{p}},k}$ per channel use are given by
\begin{equation}\label{3}
R_{{\mathrm{c}},k} = \text{log}_2{(1 + \Gamma_{{\mathrm{c}},k} )}, \quad
R_{{\mathrm{p}},k} = \text{log}_2{(1 + \Gamma_{{\mathrm{p}},k} )}, 
\end{equation}
where $\Gamma_{{\mathrm{c}},k}$ and $\Gamma_{{\mathrm{p}},k}$ are the respective instantaneous signal to interference plus noise ratio (SINR) of the common stream and the private stream at user-$k$ for a given channel use. They are given as
\begin{equation}\label{4}
\Gamma_{\mathrm{c},k}=\frac{| {{\bf{h}}_k^{\rm{H}}{{\bf{p}}_{\rm{c}}}} |^2}{\sum\nolimits_{j \in \mathcal{K}} {{{| {{\bf{h}}_k^{\rm{H}}{{\bf{p}}_j}} |}^2}}  + \sigma _n^2},\quad 
\Gamma_{\mathrm{p},k}=\frac{{| {{\bf{h}}_k^{\rm{H}}{{\bf{p}}_{k}}} |^2}}{\sum\nolimits_{j \in \mathcal{K}, j \neq k} {{{| {{\bf{h}}_k^{\rm{H}}{{\bf{p}}_j}} |}^2}}  + \sigma _n^2}. 
\end{equation}

After decoding $s_k$ into $\widehat W_{{\mathrm{p}},k}$, one more layer of ideal SIC is employed to re-encode, precode, and remove the intended private stream from the received signal.
User-$k$ then tries to eavesdrop the private messages $\{W_{{\mathrm{p}},j}| j\in\mathcal{K}\backslash\{k\}\}$ intended for other $K-1$ users. 
Therefore, the achievable wiretap rate of decoding the private stream $s_j$ at user-$k$ per channel use is 
\begin{equation}\label{5}
R_{j, k}=\log _{2}(1+\Gamma_{j,k}), \quad \forall j \in \mathcal{K} \backslash\{k\},
\end{equation}
where $\Gamma_{j,k}$ is the SINR of user-$j$'s private stream at user-$k$ and is given by
\begin{equation}\label{Gamma_jk}
\Gamma_{j, k}=\frac{|\mathbf{h}_{k}^{\rm H} \mathbf{p}_{j}|^{2}}{\sum\nolimits_{k' \in \mathcal{K}, k' \neq k, j}|\mathbf{h}_{k}^{\rm H} \mathbf{p}_{k'}|^{2}+ \sigma _n^2}, \quad \forall j \in \mathcal{K} \backslash\{k\}.
\end{equation}
The secrecy rate \cite{mukherjee2014PLSsurvey,LYC1978Gaussian} of user-$k$ is then defined as
\begin{equation}\label{6}
R_{{\mathrm{s}},k}=[R_{{\mathrm{p}},k}-\max\nolimits_{j \in \mathcal{K} \backslash\{k\}}\{R_{k,j}\}]^+,\quad\forall k\in\mathcal{K},
\end{equation}
where $\max_{j \in \mathcal{K} \backslash\{k\}}\{R_{k,j}\}$ is the largest wiretap rate at all potential eavesdroppers user-$j$, $\forall j \in \mathcal{K}\backslash\{k\}$ to decode the private message intended for user-$k$. The operation $[x]^+=\max\{x,0\}$.

Notably, to further guarantee that $W_{\mathrm{c}}$ is successfully decoded by all users, the actual transmission rate $R_{\mathrm{c}}$ for the common stream $s_{\mathrm{c}}$ should not exceed the minimum achievable rate of the common stream at all users, i.e., $\min_{k \in \mathcal{K}} R_{{\mathrm{c}},k}$. 
Moreover, $R_{\mathrm{c}}$ is shared by all users in the user set $\mathcal{K}$ where each user-$k$ is allocated to a portion $C_k$ of $R_{\mathrm{c}}$ corresponding to the rate of $W_{{\mathrm{c}},k}$. The common stream rate allocation $C_1,...,C_K$ satisfies $\sum_{k\in\mathcal{K}}C_k=R_{\mathrm{c}}$. 
Hence, the total achievable rate of user-$k$ is defined as $R_{k,{\rm tot}} = C_k+R_{{\mathrm{p}},k}$.

\section{Optimization Framework for Perfect CSIT}
In this section, the instantaneous WSR subject to the secrecy rate constraint of each user is maximized when perfect CSIT is available. 
Noting that the existing WMMSE-based algorithm for perfect CSIT in \cite{mao2018rsma} fails to deal with the non-convex secrecy rate constraint required by each user under our framework, we propose an SCA-based algorithm to iteratively optimize the precoder and common rate vector. 

For a given user weight vector ${\mathbf{u}}=[u_1,....,u_K]$, the RSMA-based secure precoding optimization problem subject to each user's secrecy rate demands is formulated as
\begin{subequations}\label{P1.1}
\begin{align}
\max _{\bf{c}, \bf{P}} & \sum\nolimits_{k \in \cal{K}} u_{k}\left(C_{k}+R_{{\mathrm{p}},k}\right) \tag{\ref{P1.1}}\\
\text { s.t. } & R_{{\mathrm{s}}, k} \geq R_{{\mathrm{s}},k}^{\rm{th}}, \quad\forall k \in \mathcal{K}, {\label{P0.b}}\\
& \sum\nolimits_{j \in \mathcal{K}} C_{j} \leq R_{{\mathrm{c}}, k}, \quad\forall k \in \mathcal{K}, \label{P0.c}\\
& \operatorname{tr}(\mathbf{P} \mathbf{P}^{\rm H}) \leq P_{max}, \label{P0.e}\\
& \mathbf{c} \geq \mathbf{0}, \label{P0.f}
\end{align}
\end{subequations}
where $R_{{\mathrm{s}},k}^{\rm{th}}$ is the secrecy rate threshold of each private stream. 
${\mathbf{c}} = [C_1,...,C_K]$ is the common rate allocation vector specifying the common rate allocated to each user. 
Constraint \eqref{P0.b} is the secrecy rate constraint of each user. 
Constraint \eqref{P0.c} ensures each user decodes the common stream successfully. 
Constraint \eqref{P0.e} is the transmit power constraint. 
Problem \eqref{P1.1} is intractable due to the non-convexity of the objective function and the constraints \eqref{P0.b}, \eqref{P0.c}. 
To address this, the adopted SCA-based algorithm will be explained explicitly in the following.

We notice that the non-convexity of problem \eqref{P1.1} is caused by the $\log$ form of the expressions of achievable rates. By introducing variables $\boldsymbol\alpha=\{\alpha_{{\mathrm{c}},k}, \alpha_{{\mathrm{p}},k}, \alpha_{k,j} \mid  \forall k\in\mathcal{K}, j\in\mathcal{K}\backslash\{k\}\}$ and substituting \eqref{3}, \eqref{5} into problem \eqref{P1.1}, constraint \eqref{P0.b} and \eqref{P0.c} can be rewritten as
\begin{subequations}
\begin{align}
&[\alpha_{{\mathrm{p}},k}-\alpha_{k,j}]^+\ge R_{{\mathrm{s}},k}^{\rm{th}}, \quad k\in\mathcal{K}, j\in\mathcal{K}\backslash\{k\},	\label{8a}\\
&\sum\nolimits_{j \in \mathcal{K}} C_{j} \leq \alpha_{{\mathrm{c}}, k}, \quad  \forall k \in \mathcal{K},\label{8b}\\
&1 + {\Gamma_{i,k}} \ge {2^{{\alpha _{i,k}}}}, \quad  \forall k\in\mathcal{K}, i\in\{{\mathrm{c}},{\mathrm{p}}\},	\label{8c}\\
&1 + {\Gamma_{k,j}} \le {2^{{\alpha _{k,j}}}}, \quad  \forall k\in\mathcal{K}, j\in \mathcal{K}\backslash\{k\}. \label{8d}
\end{align}
\end{subequations}
Due to the existence of fractional forms in the expressions of SINRs and exponential forms, \eqref{8c} and \eqref{8d} are still non-convex. 
To simplify the constraints, we further introduce ${\boldsymbol\rho}=\{\rho_{{\mathrm{c}},k}, \rho_{{\mathrm{p}},k}, \rho_{k,j}\mid \forall k\in\mathcal{K}, j\in\mathcal{K}\backslash\{k\}\}$ and rewrite them into 
\begin{subequations}
\begin{align}
&1 + {\rho_{i,k}} \ge {2^{{\alpha _{i,k}}}}, \quad \forall k\in\mathcal{K}, i\in\{{\mathrm{c}},{\mathrm{p}}\},	\label{8c2}\\
&1 + {\rho_{k,j}} \le {2^{{\alpha _{k,j}}}}, \quad \forall k\in\mathcal{K},\forall j\in \mathcal{K}\backslash\{k\}, \label{8d2}
\end{align}
\end{subequations}
and 
\begin{subequations}
\begin{align}
&{\Gamma _{{\mathrm{p}},k}} \ge {\rho _{{\mathrm{p}},k}}, \quad \forall k\in\mathcal{K}, \label{8e2}\\
&{\Gamma _{k,j}} \le {\rho _{k,j}}, \quad \forall k\in\mathcal{K},\forall j\in \mathcal{K}\backslash\{k\}.\label{8f2}
\end{align}
\end{subequations}
Plugging \eqref{4}, \eqref{Gamma_jk} into \eqref{8e2}, \eqref{8f2}, the problem \eqref{P1.1} is rewritten as
\begin{subequations}\label{P1.2}
\begin{align}
\max _{\mathbf{c}, \mathbf{P}, {\boldsymbol\alpha}, {\boldsymbol\rho}} & \sum\nolimits_{k \in \mathcal{K}} u_{k}\big(C_{k}+\alpha_{{\mathrm{p}}, k}\big) \tag{\ref{P1.2}}\\
& \frac{{{{\big| {{\bf{h}}_k^{\rm H}{{\bf{p}}_{\mathrm{c}}}}\big|}^2}}}{{\sum\nolimits_{j \in \mathcal{K}} {{{\big| {{\bf{h}}_k^{\rm H}{{\bf{p}}_j}}\big|}^2}}  + \sigma_n^2}} \ge {\rho _{{\mathrm{c}},k}},\quad \forall k\in\mathcal{K},\label{9c}\\
& \frac{{{{\big| {{\bf{h}}_k^{\rm H}{{\bf{p}}_k}} \big|}^2}}}{{\sum\nolimits_{j \in \mathcal{K}, j \ne k} {{{\big| {{\bf{h}}_k^{\rm H}{{\bf{p}}_j}} \big|}^2}}  + \sigma_n^2}} \ge {\rho _{{\mathrm{p}},k}},\quad \forall k\in\mathcal{K},\label{9d}\\
& \frac{{{{\big| {{\bf{h}}_j^{\rm H}{{\bf{p}}_k}} \big|}^2}}}{{\sum\nolimits_{k' \in \mathcal{K}, k' \ne k,j} {{{\big| {{\bf{h}}_j^{\rm H}{{\bf{p}}_{k'}}} \big|}^2}}  + \sigma_n^2}} \le {\rho _{k,j}}, \quad \forall j \in \mathcal{K}\backslash\{k\}, \label{9e}\\
&\nonumber \eqref{P0.e}, \eqref{P0.f}, \eqref{8a}, \eqref{8b}, \eqref{8c2}, \eqref{8d2}. 
\end{align}
\end{subequations}

However, problem \eqref{P1.2} is still non-convex due to constraints \eqref{8d2} and \eqref{9c}-\eqref{9e}. 
To address this, we adopt the first-order Taylor expansion \cite{boyd2004convex} to linearly approximate \eqref{8d2} as
\begin{equation}\label{10}
1+\rho_{k,j}\le 2^{\alpha_{k,j}^{[n]}}[1+\ln2(\alpha_{k,j}-\alpha_{k,j}^{[n]})],\quad\forall k \in \mathcal{K},\forall j\in\mathcal{K}\backslash\{k\},
\end{equation}
where $\alpha_{k,j}^{[n]}$ denotes the optimized $\alpha_{k,j}$ obtained from the $[n]$-th iteration. 
Additionally, to reformulate constraints \eqref{9c} and \eqref{9d}, we introduce another vector $\boldsymbol\beta=\{\beta_{{\mathrm{c}},k},\beta_{{\mathrm{p}},k}\mid  \forall k\in\mathcal{K}\}$ to replace the denominator of the inequalities, hence \eqref{9c} and \eqref{9d} are equivalent to
\begin{subequations}
\begin{align}
&\frac{{{{\big| {{\bf{h}}_k^{\rm H}{{\bf{p}}_i}} \big|}^2}}}{{{\beta _{i,k}}}} \ge {\rho _{i,k}}, \quad \forall k \in {\cal K}, i\in\{{\mathrm{c}},{\mathrm{p}}\},	\label{11a}\\
&\sum\nolimits_{j \in {\cal K}} {{{\big| {{\bf{h}}_k^{\rm H}{{\bf{p}}_j}} \big|}^2}}  + \sigma_n^2 \le {\beta _{{\mathrm{c}},k}},	\label{11b}\\
&\sum\nolimits_{j \in \mathcal{K}, j \ne k} {{{\big| {{\bf{h}}_k^{\rm H}{{\bf{p}}_j}} \big|}^2}}  + \sigma_n^2 \le {\beta _{{\mathrm{p}},k}}.	\label{11c}
\end{align}
\end{subequations}
Similarly, the non-convex constraints \eqref{11a} and \eqref{9e} can be approximately reconstructed into convex ones by adopting the first-order Taylor expansion, which are given by
\begin{equation}\label{12}
\begin{aligned}
\frac{2 \Re\Big\{\big(\mathbf{p}_{i}^{[n]}\big)^{\rm H} \mathbf{h}_{k} \mathbf{h}_{k}^{\rm H} \mathbf{p}_{i}\Big\}}{\beta_{i,k}^{[n]}} 
-\frac{\big|\mathbf{h}_{k}^{\rm H} \mathbf{p}_{i}^{[n]}\big|^{2} \beta_{i,k}}{\big(\beta_{i,k}^{[n]}\big)^{2}} \geq \rho_{i,k}, \quad\forall k \in \mathcal{K}, i\in\{{\mathrm{c}},{\mathrm{p}}\},
\end{aligned}
\end{equation}
and
\begin{equation}\label{13}
\rho_{k,j}^{[n]}\sum\limits_{\substack {k' \in {\cal K} \\ k' \ne k,j}}  \Big({{\mathop{2\Re}\nolimits} \bigg( {{{\big( {{\bf{p}}_{_{k'}}^{[n]}} \big)}^{\rm H}}{{\bf{h}}_j}{\bf{h}}_j^{\rm H}{{\bf{p}}_{k'}}} \bigg)} - {{{\big| {{\bf{h}}_j^{\rm H}{\bf{p}}_{_{k'}}^{[n]}} \big|}^2}} \Big)
+\rho_{k,j}\bigg( {\sum\limits_{\substack {k' \in {\cal K} \\ k' \ne k,j}} {{{\big| {{\bf{h}}_j^{\rm H}{\bf{p}}_{_{k'}}^{[n]}} \big|}^2}}  + \sigma_n^2} \bigg) \ge {\big| {{\bf{h}}_j^{\rm H}{{\bf{p}}_k}} \big|^2},
\end{equation}
where $\mathbf{p}_{i}^{[n]}$, ${{\bf{p}}_{_{k'}}^{[n]}}$, ${\beta_{i,k}^{[n]}}$ and $\rho_{k,j}^{[n]}$ are the respective optimized precoders $\mathbf{p}_{i}$, ${\bf{p}}_{k'}$, and $\beta_{i,k}$, $\rho_{k,j}$ obtained in the $[n]$-th iteration.

Consequently, based on \eqref{10}, \eqref{12}, \eqref{13}, using the solution $({\mathbf P}^{[n-1]}, {\boldsymbol\alpha}^{[n-1]}, {\boldsymbol\beta}^{[n-1]}, {\boldsymbol\rho}^{[n-1]})$ obtained from the $[n-1]$-th iteration, we solve the following problem at the $[n]$-th iteration:
\begin{subequations}\label{P1.3}
\begin{align}
\max _{\mathbf{c}, \mathbf{P}, \boldsymbol{\alpha}, \boldsymbol{\beta}, \boldsymbol{\rho}} & \sum\nolimits_{k \in \mathcal{K}} u_{k}\left(C_{k}+\alpha_{{\mathrm{p}}, k}\right) \tag{\ref{P1.3}}\\
\text { s.t. } &\nonumber\eqref{8a}, \eqref{8b}, \eqref{8c2}, \eqref{10}, \eqref{11b}, \eqref{11c}, \eqref{12}, \eqref{13}, \eqref{P0.e}, \eqref{P0.f}.
\end{align}
\end{subequations}
The transformed problem \eqref{P1.3} is now convex and can be effectively solved by applying state-of-art toolboxes, such as CVX. 
The proposed SCA-based algorithm is summarized in Algorithm 1. 
At each iteration, the optimized parameters $({\mathbf P}^{[n-1]},{\boldsymbol\alpha}^{[n-1]}, {\boldsymbol\beta}^{[n-1]}, {\boldsymbol\rho}^{[n-1]})$ obtained from last iteration serve as inputs for optimizing problem \eqref{P1.3} until the difference of WSR between two successive iterations satisfies a preset threshold.

\emph{Convergence analysis:} 
As \eqref{10}, \eqref{12} and \eqref{13} are the lower bound approximations of \eqref{8d2}, \eqref{11a} and \eqref{9e}, the optimal solution obtained in the $[n-1]$-th iteration also serves as a feasible solution at the $[n]$-th iteration. Therefore, the corresponding optimized objective value of the $[n-1]$-th iteration is no larger than that of the $[n]$-th iteration. Besides, the objective function in problem \eqref{P1.3} is bounded by the transmit power constraint \eqref{P0.e}, the convergence is hence guaranteed. However, it should be noted that there is no guarantee of global optimality since only the KKT conditions of the original problem could be ensured after the termination of the iteration process. 
\begin{algorithm}[!t]
\caption{SCA-based algorithm for perfect CSIT setup}
\label{Algorithm 1}
\begin{algorithmic}[1]
    \Require
    the tolerance $\epsilon$, the secrecy threshold $R_{{\mathrm{s}},k}^{\rm{th}}$, the transmit power limit $P_{max}$.
    \Ensure
    the obtained WSR, the optimal secure precoder ${\mathbf P}^{\rm opt}$ and common rate vector ${\mathbf c}^{\rm opt}$. 
    \State Set threshold $\epsilon $, set iteration index $n:=0$;\\
     Initialize $({\mathbf P}^{[0]},{\boldsymbol\alpha}^{[0]}, {\boldsymbol\beta}^{[0]})$ and the obtained ${\rm WSR}^{[0]}$;
    \Repeat:
    \State Using $({\mathbf P}^{[n-1]},{\boldsymbol\alpha}^{[n-1]}, {\boldsymbol\beta}^{[n-1]})$ obtained from last iteration, solve problem $\rm P1.2$ and assign
    \State  the obtained optimized variables $({\mathbf P}^{\rm *},{\boldsymbol\alpha}^{\rm *}, {\boldsymbol\beta}^{\rm *})$ as $({\mathbf P}^{[n]},{\boldsymbol\alpha}^{[n]}, {\boldsymbol\beta}^{[n]})$ and optimized objective  
    \State value ${\rm WSR}^{*}$ as ${\rm WSR}^{[n]}$; 
    \State Update iteration: $n\leftarrow n+1$;
    \Until \textit{Stopping criterion satisfying:}
    \begin{equation}
    \nonumber
    \left| {\rm WSR}^{[n]}-{\rm WSR}^{[n-1]} \right|\le \epsilon;
    \end{equation}
    \State \Return ${\mathbf P}^{\rm opt}={\mathbf P}^{[n]}$, ${\mathbf c}^{\rm opt}={\mathbf c}^{[n]}$.
\end{algorithmic}
\end{algorithm}

\section{Optimization Framework for imperfect CSIT}
In this section, we consider a more practical case when only imperfect CSIT is available. 
By extending the WMMSE algorithm in \cite{Hamdi2016RSWMMSE}, we propose a joint WMMSE and SCA based AO algorithm. In our algorithm, the relationship between WSR and WMMSE is first established to simplify the original problem. 
To deal with the difference-of-convex structure introduced by the secrecy rate constraint in the transformed non-convex problem, the SCA-based approach is applied. 
The WESR is then maximized by optimizing the precoder and the common rate vector.

\subsection{Problem Reformulation}
We first discuss the channel model where only partial CSIT, denoted by $\widehat{\mathbf{H}} \triangleq[\widehat{\mathbf{h}}_{1}, \ldots, \widehat{\mathbf{h}}_{K}]$, is available. 
The estimated channel error is denoted by $\widetilde{\mathbf{H}} \triangleq[\widetilde{\mathbf{h}}_{1}, \ldots, \widetilde{\mathbf{h}}_{K}]$. 
Each element of its column vector follows the independent and identical zero-mean complex Gaussian distribution, i.e., $\mathrm{E}\{\widetilde{\mathbf{h}}_{k} \widetilde{\mathbf{h}}_{k}^{\rm H}\}=\sigma_{{\mathrm{e}},k}^2\bf I$. 
$\sigma_{{\mathrm{e}},k}^2 \sim O(P_{max}^{-\delta})$ and $\delta\in[0,\infty)$ is the scaling factor characterizing the variance of errors and the quality of CSIT in the high SNR regime, respectively. Denoting ${\mathbf{H}} \triangleq[{\mathbf{h}}_{1}, \ldots, {\mathbf{h}}_{K}]$, the imperfect CSIT model is \cite{Hamdi2016RSWMMSE} 
\begin{equation}\label{H_relationship}
\bf H=\widehat{\bf H}+\tilde{\bf H}.
\end{equation}

Maximizing the instantaneous WSR solely based on the estimated channel state $\widehat{\bf H}$ at the transmitter may lead to transmission at undecodable rates since it neglects the introduced errors caused by MUI. 
Therefore, when CSIT is imperfect, we aim at maximizing the WESR to characterize a long-term WSR performance of all users. 
The WESR is calculated as a weighted sum of all users' ergodic rate, defined as ${\rm WESR}({\bf H, \widehat{\bf H}}) \triangleq \sum_{k\in\mathcal{K}} u_k  \{\mathrm{E}_{\{{\bf H, \widehat{\bf H}}\}}\{R_{{\mathrm{p}},k} + C_k\}\}$.  
For $\forall k \in \mathcal{K}$, $j\in\mathcal{K}\backslash\{k\}$, the WESR maximization problem is formulated as:
\begin{subequations}\label{WESR_problem}
\begin{align}
\max_{{\bf c}, {\bf P}} &\sum\nolimits_{k\in\mathcal{K}} u_k  \{\mathrm{E}_{\{{\bf H, \widehat{\bf H}}\}}\{R_{{\mathrm{p}},k} + C_k\}\} \tag{\ref{WESR_problem}}\\
\text { s.t. } & \mathrm{E}_{\{{\bf H, \widehat{\bf H}}\}}\{R_{{\mathrm{p}},k}\}-\mathrm{E}_{\{{\bf H, \widehat{\bf H}}\}}\{R_{k,j}\} \geq R_{{\mathrm{s}},k}^{\rm th}, \label{P2.1.b} \\
& \sum\nolimits_{j \in \mathcal{K}} \mathrm{E}_{\{{\bf H, \widehat{\bf H}}\}} \{C_j\}  \leq \mathrm{E}_{\{{\bf H, \widehat{\bf H}}\}}\{R_{{\mathrm{c}},k}\}, \label{P2.1.c}\\
& \mathrm{E}_{\{{\bf H}, \widehat{\bf H}\}} \{C_{j}\} \geq \mathbf{0}, \forall j\in\mathcal{K}, \label{P2.1.e}\\
& \mathrm{E}_{\{{\bf H, \widehat{\bf H}}\}}\{\operatorname{tr}(\mathbf{P} \mathbf{P}^{\rm H})\}\leq P_{max},
\end{align}
\end{subequations}
where constraint \eqref{P2.1.b} ensures that the ergodic secrecy rate of each private stream satisfies the corresponding security requirement $R_{{\rm s},k}^{\rm th}$. Constraint \eqref{P2.1.c} ensures the ergodic common rate can be successfully decoded by all receivers.

However, problem \eqref{WESR_problem} is intractable since the joint probability density distribution $f_{\bf H,\widehat{\bf H}}({\bf H, \widehat{\bf H}})$ is unknown to the transmitter. 
Only the knowledge of the conditional density $f_{\bf H|\widehat{\bf H}}({\bf H|\widehat{\bf H}})$ is available. 
We know from the law of total expectation that, the ergodic rate (ER) performance over all possible channel uses can be characterized by averaging the averaged rates (AR), each of which measures the short-term expected rate performance over the CSIT error distribution for a given channel estimate $\widehat{\bf H}$. 
The relationship between ER and AR is demonstrated in \eqref{ER_rewrite} \cite{Hamdi2016RSWMMSE}, where the AR under a given channel estimate $\widehat{\bf H}$ is defined as $\bar{R}_{k}(\widehat{\bf H}) \triangleq \mathrm{E}_{\mathrm{H} \mid \widehat {\mathrm{H}}}\{R_{k} \mid \widehat{\mathbf{H}}\}$.
\begin{equation}\label{ER_rewrite}
\begin{aligned}
\mathrm{E}_{\{{\bf H, \widehat{\bf H}}\}}\big\{R_{k}(\mathbf{H}, \widehat{\mathbf{H}})\big\} &=\mathrm{E}_{\widehat{\bf {H}}}\bigg\{\mathrm{E}_{\bf{H} \mid \widehat{\bf{H}}}\big\{R_{k}(\mathbf{H}, \widehat{\mathbf{H}}) \mid \widehat{\mathbf{H}}\big\}\bigg\}
=\mathrm{E}_{\widehat{\bf{H}}}\big\{\bar{R}_{k}(\widehat{\mathbf{H}})\big\}.
\end{aligned}
\end{equation}

Hence, under imperfect CSIT, problem \eqref{WESR_problem} 
can be equivalently decomposed into a two-step optimization problem, i.e., to first maximize the weighted average sum-rate (WASR) under a given channel estimate $\widehat{\bf H}$, followed by averaging WASR over all possible channel estimates.
Accordingly, We define the achievable common AR at user-$k$ as $\bar{R}_{{\mathrm{c}},k}(\widehat{\bf H}) \triangleq \mathrm{E}_{\mathrm{H} | \widehat {\mathrm{H}}}\{R_{{\mathrm{c}},k} | \widehat{\mathbf{H}}\}$, the achievable private AR at user-$k$ as $\bar{R}_{{\mathrm{p}},k}(\widehat{\bf H}) \triangleq \mathrm{E}_{\mathrm{H} | \widehat {\mathrm{H}}}\{R_{{\mathrm{p}},k} | \widehat{\mathbf{H}}\}$ and the achievable AR of user-$j$ eavesdropping the private stream of user-$k$ as $\bar{R}_{k,j}(\widehat{\bf H}) \triangleq \mathrm{E}_{\mathrm{H} | \widehat {\mathrm{H}}}\{R_{k,j} | \widehat{\mathbf{H}}\}$. 
The WASR maximization problem under a given channel estimate $\widehat{\bf H}$ is then formulated as
\begin{subequations}\label{P2.1_ASRform}
\begin{align}
\max _{\mathbf{\bar c}, \mathbf P} & \sum\nolimits_{k \in \mathcal{K}} u_{k}\big(\bar C_{k}+\bar R_{{\mathrm{p}},k}(\widehat{\bf H})\big) \tag{\ref{P2.1_ASRform}}\\
\text { s.t. } & \bar R_{{\mathrm{p}},k}(\widehat{\bf H})-\bar R_{k,j}(\widehat{\bf H}) \geq R_{{\mathrm{s}},k}^{\rm th}, \quad \forall k \in \mathcal{K}, \\
& \sum\nolimits_{j \in \mathcal{K}} \bar C_{j} \leq \bar R_{{\mathrm{c}}, k}(\widehat{\bf H}), \quad \forall k \in \mathcal{K}, \\
& \mathbf{\bar c} \geq \mathbf{0}, \label{P2.1.e}\\
& \nonumber\eqref{P0.e},
\end{align}
\end{subequations}
where 
$\bar{\bf c}=[\bar C_1,...,\bar C_K]$ is the optimized common AR vector specifying common AR allocated to each user.

Clearly, problem \eqref{P2.1_ASRform} is a stochastic non-convex optimization problem since $\bar R_{{\mathrm{c}},k}$, $\bar R_{{\mathrm{p}},k}$ and $\bar R_{k,j}$ are expectations over the random variable matrix of the estimated channel error $\tilde{\bf H}$. To address problem \eqref{P2.1_ASRform}, we first adopt SAA approach to handle its stochastic nature \cite{Hamdi2016RSWMMSE}. 
After approximating the stochastic ARs into the corresponding deterministic expressions, a dual-loop AO approach is proposed to iteratively find the solutions, where a WMMSE method is in the outer loop to optimize the MMSE parameters and an SCA-based algorithm is in the inner loop to optimize the precoder and common rate vector.
   
\subsection{Sample Average Approximation Approach}

The stochastic feature of \eqref{P2.1_ASRform} is handled using the SAA approach \cite{Hamdi2016RSWMMSE}. 
At the transmitter, with known conditional distribution density $f_{\bf H| \widehat{\bf H}}(\bf H| \widehat{\bf H})$, the set of sampled channel realizations $\mathbb{H}^{(M)}$ is derived under each given channel estimate $\widehat {\bf H}$, i.e., $\mathbb{H}^{(M)} \triangleq\left\{\mathbf{H}^{(m)}=\widehat{\mathbf{H}}+\widetilde{\mathbf{H}}^{(m)} \mid \widehat{\mathbf{H}}, m \in \mathcal{M}\right\}$,  where $\cal{M}$ is the index set of user channel samples, ${\cal{M}}=\{1,...m,...,M\}$. 
The sample average functions (SAFs) is adopted to approximate the ARs, which are defined as
\begin{equation}\label{SAA_def}
\begin{aligned}
\bar{R}_{i, k}^{(M)}(\widehat{\bf H}) &\triangleq \frac{1}{M} \sum_{m=1}^{M} R_{i, k}^{(m)} (\widehat{\bf H}), \quad i\in\{\mathrm{c},\mathrm{p}\},\\
\bar{R}_{k,j}^{(M)}(\widehat{\bf H}) &\triangleq \frac{1}{M} \sum_{m=1}^{M} R_{k,j}^{(m)} (\widehat{\bf H}), \quad j\in\mathcal{K}\backslash\{k\},
\end{aligned}
\end{equation}
where the achievable rate samples $R_{i,k}^{(m)}$ and $R_{k,j}^{(m)}$ under a given channel estimate $\widehat{\bf H}$ are defined as the function of a channel realization ${\bf H}^{(m)}$: $R_{i, k}^{(m)}(\widehat{\bf H})=R_{i, k}({\bf H}^{(m)},\widehat{\bf H})$, $R_{k,j}^{(m)}(\widehat{\bf H})=R_{k,j}({\bf H}^{(m)},\widehat{\bf H})$. 

By the strong law of large numbers (LLN), with the number of channel samples approaching infinity, the ARs with a given channel estimate $\widehat {\bf H}$ can be equivalently expressed by their corresponding SAFs:
\begin{equation}\label{LNN_R}
\begin{aligned}
\bar R_{i,k} (\widehat{\bf {H}})&=\lim_{M \to \infty} \bar R^{(M)}(\widehat{\bf {H}}), \quad i\in\{\mathrm{c},\mathrm{p}\},\\
\bar R_{k,j} (\widehat{\bf {H}})&=\lim_{M \to \infty} \bar R^{(M)}(\widehat{\bf {H}}), \quad j\in \mathcal{K}\backslash\{k\}. 
\end{aligned}
\end{equation}
Consequently, plugging \eqref{SAA_def}, the stochastic problem \eqref{P2.1_ASRform} is reformulated to a deterministic form:
\begin{subequations}\label{P2.2_SAA}
\begin{align}
\max _{\mathbf{\bar c}, \mathbf{P}} & \sum\nolimits_{k \in \mathcal{K}} u_{k}(\bar C_{k}+\bar R_{\mathrm{p},k}^{(M)}(\widehat{\bf H})) \tag{\ref{P2.2_SAA}}\\
\text { s.t. } & \bar R_{\mathrm{p},k}^{(M)}(\widehat{\bf H})-\bar R_{k,j}^{(M)} (\widehat{\bf H})\geq R_{\mathrm{s},k}^{\rm th}, \quad \forall k \in \mathcal{K}, {\label{P2.2.b}}\\
& \sum\nolimits_{j \in \mathcal{K}} \bar C_{j} \leq \bar R_{\mathrm{c}, k}^{(M)}(\widehat{\bf H}), \quad \forall k \in \mathcal{K}, \label{P2.2.c}\\
& \nonumber \eqref{P0.e}, \eqref{P2.1.e}.
\end{align}
\end{subequations}

\subsection{Joint WMMSE and SCA based AO algorithm}
Problem \eqref{P2.2_SAA} is still intractable as both the objective function and the constraints \eqref{P2.2.b}, \eqref{P2.2.c} are non-convex. 
To address this, we first apply the WMMSE algorithm specified in \cite{Hamdi2016RSWMMSE} to rewrite the non-convex expression of $\bar R_{\mathrm{p}, k}^{(M)}$, $\bar R_{\mathrm{c}, k}^{(M)}$ and $\bar R_{k,j}^{(M)}$ by their corresponding equivalent forms and reformulate problem \eqref{P2.2_SAA}. 
Then, an SCA-based algorithm is proposed to deal with the non-convex constraint in the reformulated problem.

Under the RS framework, the common stream is first decoded at user-$k$ by applying a common-stream equalizer $g_{\mathrm{c},k}$ and the estimated common stream is given by $\widehat{s}_{\mathrm{c},k}=g_{\mathrm{c},k}y_k$. 
After subtracting the received common part, the estimate of the private stream for user-$k$ is obtained by applying a private-stream equalizer $g_{\mathrm{p},k}$ to the remaining part, which is denoted by $\widehat{s_k}=g_{\mathrm{p},k}(y_k-{\bf h}_k^{\rm H}{\bf p}_{\mathrm{c}} {\widehat s_{\mathrm{c},k}})$. 
After that, each user tries to eavesdrop the private stream intended to other users using a similar method. 
Therefore, the estimation of the private stream $s_j$ at user-$k$ is performed as $\widehat{s}_{j,k}=g_{j,k}(y_k-{\bf h}_k^{\rm H}{\bf p}_c {\widehat s_{\mathrm{c},k}}-{\bf h}_k^{\rm H}{\bf p}_k {\widehat s_{k}})$, where $g_{j,k}$ is the equalizer at user-$k$ to wiretap the private stream for user-$j$. The estimation errors for decoding $s_{\mathrm{c}}$, $s_k$ and $s_{j,k}$ are
\begin{equation}\label{epsilon_def}
\begin{aligned}
\varepsilon_{\mathrm{c}, k} &\triangleq \mathrm{E}\{|\widehat{s}_{\mathrm{c}, k}-s_{\mathrm{c}}|^{2}\}
=\left|g_{\mathrm{c}, k}\right|^{2} T_{\mathrm{c}, k}-2 \Re\left\{g_{\mathrm{c}, k} \mathbf{h}_{k}^{\rm H} \mathbf{p}_{\mathrm{c}}\right\}+1, \\
\varepsilon_{\mathrm{p},k} &\triangleq \mathrm{E}\{|\widehat{s}_{k}-s_{k}|^{2}\}=\left|g_{\mathrm{p},k}\right|^{2} T_{\mathrm{p},k}-2 \Re\left\{g_{\mathrm{p},k} \mathbf{h}_{k}^{\rm H} \mathbf{p}_{k}\right\}+1,\\
\varepsilon_{j,k} &\triangleq \mathrm{E}\{|\widehat{s}_{j,k}-s_{j}|^{2}\}=\left|g_{j,k}\right|^{2} T_{j,k}-2 \Re\left\{g_{j,k} \mathbf{h}_{k}^{\rm H} \mathbf{p}_{j}\right\}+1,
\end{aligned}
\end{equation}
where 
\begin{equation}\label{T_def}
\begin{aligned}
T_{\mathrm{c}, k} &=|\mathbf{h}_{k}^{\rm H} \mathbf{p}_{\mathrm{c}}|^{2}+\sum_{k' \in \cal{K}}|\mathbf{h}_{k}^{\rm H} \mathbf{p}_{k'}|^{2}+\sigma_{\mathrm{n}}^{2},\\
T_{\mathrm{p},k} &= I_{\mathrm{c}, k}=T_{\mathrm{c}, k}-|\mathbf{h}_{k}^{\rm H} \mathbf{p}_{\mathrm{c}}|^{2},\\
T_{j,k} &=I_{{\rm p}, k} =T_{\mathrm{p},k}-|\mathbf{h}_{k}^{\rm H} \mathbf{p}_{k}|^{2},\\
I_{j,k} &=T_{j,k}-|{\bf h}_k^{\rm H}{\bf p}_j|^2.
\end{aligned}
\end{equation}
By minimizing the estimation errors in \eqref{epsilon_def}, the optimal MMSE equalizers at user-$k$ are given by
\begin{equation}\label{g_def}
\begin{aligned}
g_{i, k}^{\text {MMSE}}&=\mathbf{p}_{i}^{\text H} \mathbf{h}_{k} T_{i, k}^{-1}, \quad i\in\{\mathrm{c},j\mid j\in\mathcal{K}\backslash\{k\}\},\qquad
g_{\mathrm{p},k}^{\text{MMSE}}&=\mathbf{p}_{k}^{\text H} \mathbf{h}_{k} T_{\mathrm{p},k}^{-1}.
\end{aligned}
\end{equation}
Plugging \eqref{g_def} back into \eqref{epsilon_def} yields
\begin{equation}\label{epsilon_opt}
\varepsilon_{i, k}^{\text{MMSE}}  \triangleq \min _{{g}_{i, k}} \varepsilon_{i, k}=T_{i, k}^{-1} I_{i, k},\quad i\in\{\mathrm{c},\mathrm{p},j\mid j\in\mathcal{K}\backslash\{k\}\}.
\end{equation}

To establish the relationship between rate and WMMSE, the weighted mean square errors (WMSE) are defined as
\begin{equation}\label{xi_def}
\xi_{i, k}=\omega_{i, k} \varepsilon_{i, k}-\log _{2}(\omega_{i, k}), \quad i\in\{\mathrm{c},\mathrm{p},j\mid j\in\mathcal{K}\backslash\{k\}\},
\end{equation} 
where $\omega_{\mathrm{c},k}$, $\omega_{\mathrm{p},k}$ and $\omega_{j,k}$ are the MMSE weights for decoding $s_{\mathrm{c},k}$, $s_k$ and $s_{j,k}$ at user-$k$, respectively. 
Substituting \eqref{epsilon_def} into \eqref{xi_def}, the relationship between WMSEs and MMSE equalizers, MMSE weights and precoders is derived as shown in \eqref{xi_c_trans}-\eqref{xi_kj_trans}. 
\newcounter{TempEqCnt}
\setcounter{TempEqCnt}{\value{equation}}
\setcounter{equation}{28}
\begin{figure*}[ht]
\begin{equation}\label{xi_c_trans}
\xi_{{\mathrm c},k} = \omega_{{\mathrm c},k}{|{g_{{\mathrm c},k}}|^2}\Big( {|{\bf h}_k^{\rm H}{{\bf p}_{\mathrm c}}|^2 + \sum\limits_{k' \in \cal {K}} |{\bf{h}}_k^{\rm H}{{\bf{p}}_{k'}}|^2 + \sigma _n^2} \Big) - 2\Re \{ {\omega_{{\mathrm c},k} g_{{\mathrm c},k} {\bf h}_k^{\rm H} {\bf p}_{\mathrm c}} \} + \omega_{{\mathrm c},k} - \log_2({\omega_{{\mathrm c},k}}),
\end{equation}
\begin{equation}\label{xi_k_trans}
\xi_{{\mathrm p},k} = \omega_{{\mathrm p},k}{|{g_{{\mathrm p},k}}|^2}\Big( { \sum\limits_{k' \in \cal {K}} |{\bf{h}}_k^{\rm H}{{\bf{p}}_{k'}}|^2 + \sigma _n^2} \Big) - 2\Re \{ {\omega_{{\mathrm p},k} g_{{\mathrm p},k} {\bf h}_k^{\rm H} {\bf p}_k} \} + \omega_{{\mathrm p},k} - \log_2({\omega_{{\mathrm p},k}}),
\end{equation}
\begin{equation}\label{xi_kj_trans}
\xi _{j,k} = \omega_{j,k}{|{g_{j,k}}|^2}\Big( { \sum\limits_{k' \neq k} |{\bf{h}}_k^{\rm H}{{\bf{p}}_{k'}}|^2 + \sigma _n^2} \Big) - 2\Re \{ {\omega_{j,k} g_{j,k} {\bf h}_k^{\rm H} {\bf p}_j} \} + \omega_{j,k} - \log_2({\omega_{j,k}}).
\end{equation}
\end{figure*}

Therefore, with given MMSE equalizers and MMSE weights, the WMSEs are convex with respect to precoders $\mathbf{p}_{\mathrm c}$ and $\mathbf{p}_{k}$, respectively.
Optimizing WMSEs with respect to MMSE weights and MMSE equalizers yields
\begin{align}
\omega_{i,k}^{*}&=\omega_{i,k}^{\text{MMSE}} \triangleq\left(\varepsilon_{i,k}^{\text{MMSE}}\right)^{-1},\label{W_def}\\
g_{i,k}^*&=g_{i,k}^{\text{MMSE}}, \quad i\in\{\mathrm{c},\mathrm{p},j\mid j\in\mathcal{K}\backslash\{k\}\}. \label{g_opt}
\end{align}
Taking \eqref{W_def} and \eqref{g_opt} back to \eqref{xi_def}, we finally have the Rate-WMMSE relationship
\begin{equation}\label{Rate-WMMSE}
\xi_{i, k}^{\text{MMSE}}({\bf H}^{(m)},\widehat{\bf H})  \triangleq \min_{\omega_{i,k}, g_{i,k}} \xi_{i,k}=1-R_{i,k}({\bf H}^{(m)},\widehat{\bf H}), \quad i\in\{\mathrm{c},\mathrm{p},j| j\in\mathcal{K}\backslash\{k\}\}.
\end{equation}
The SAFs of the WMSEs are obtained by averaging over $M$ channel samples, i.e., $\bar \xi_{i,k}^{(M)}=\frac{1}{M}\sum_{m=1}^{M} \xi_{i,k}({\bf H}^{(m)},\widehat{\bf H})$, $i\in\{{\mathrm c}, {\mathrm p}\}$ and $\bar \xi_{k,j}^{(M)}=\frac{1}{M}\sum_{m=1}^{M} \xi_{k,j}({\bf H}^{(m)},\widehat{\bf H})$. 
Thus, when $M\to\infty$,
\begin{equation}\label{Ergodic_rw}
\begin{aligned}
\bar \xi_{i,k}^{\text {MMSE}}&=\min_{\omega_{i,k},g_{i,k}}\bar \xi_{i,k}^{(M)}=1-\bar R_{i,k}^{(M)}, \quad i\in\{{\rm c}, {\rm p}\}\\
\bar \xi_{k,j}^{\text {MMSE}}&=\min_{\omega_{k,j},g_{k,j}}\bar \xi_{k,j}^{(M)}=1-\bar R_{k,j}^{(M)}, \quad j\in\mathcal{K}\backslash\{k\}.
\end{aligned}
\end{equation}
Then, applying \eqref{Ergodic_rw}, problem \eqref{P2.2_SAA} is equivalently rewritten as
\begin{subequations}\label{P1}
\begin{align}
\min _{\mathbf{\bar x}, \mathbf{P}, \mathbf{\bf \Omega}, \mathbf{G}} & \sum\nolimits_{k \in \mathcal{K}} u_{k} (\bar{\xi}_{{\mathrm p},k}^{(M)}+\bar X_{k} ) \tag{\ref{P1}}\\
\text { s.t. } & \bar{\xi}_{{\mathrm p},k}^{(M)}-\bar{\xi}_{k,j}^{(M)} \leq -R_{{\mathrm s},k}^{\rm th}, \forall k \in \mathcal{K}, {\label{P2.3.b}}\\
& -\sum\nolimits_{i \in \mathcal{K}} \bar X_{i} +\bar{\xi}_{{\mathrm c},k}^{(M)} \leq 1, \forall k \in \mathcal{K}, \label {P2.3.c}\\
& \mathbf{\bar x} \leq \mathbf{0}, 
\label{P2.3.e}\\
& \nonumber \eqref{P0.e},
\end{align}
\end{subequations}
where ${\bf {\bar x}}=-{\bf {\bar c}}=[\bar X_1,...,\bar X_K]$, ${\bf \Omega}=\{\omega_{i,k},\omega_{k,j},\mid i\in\{\mathrm{c},\mathrm{p}\}, k\in\mathcal{K}, j\in\mathcal{K}\backslash\{k\}\}$ is the vector consisting of optimal MMSE weights, and ${\bf G}=\{g_{i,k},g_{k,j}\mid i\in\{\mathrm{c},\mathrm{p}\}, k\in\mathcal{K}, j\in\mathcal{K}\backslash\{k\}\}$ is the vector consisting of optimal MMSE equalizers.  The equivalence proof between \eqref{P1} and \eqref{P2.2_SAA} is given in Appendix A.
To decouple the variables $\bf {\bar x}$, $\bf P$ from MMSE equalizers $\bf G$ and weights $\bf \Omega$, we adopt AO approach to iteratively optimize $\bf \Omega$ and $\bf G$ in the outer loop and the secure precoder and common AR vector in the inner loop. 
To be specific, for all common and private stream related parameters, we first define, $\forall i\in\{\mathrm c,\mathrm p\}$,
\begin{equation}\label{vars_def}
\begin{aligned}
\bar t_{i,k} &\triangleq \frac{1}{M} \sum_{m=1}^{M}t_{i,k}^{(m)} =\frac{1}{M} \sum_{m=1}^{M}\big(\omega_{i,k}^{(m)}|g_{i,k}^{(m)}|^{2}\big), \\
\bar \Psi_{i,k} &\triangleq \frac{1}{M} \sum_{m=1}^{M} \Psi_{i,k}^{(m)} = \frac{1}{M} \sum_{m=1}^{M} \big(t_{i,k}^{(m)} {\mathbf h}_{k}^{(m)} {\mathbf h}_{k}^{(m)H}\big), \\
\bar {\mathbf{f}}_{i,k} &\triangleq \frac{1}{M} \sum_{m=1}^{M} \mathbf{f}_{i,k}^{(m)} =\frac{1}{M} \sum_{m=1}^{M} \big(\omega_{i,k}^{(m)} \mathbf{h}_{k}^{(m)} g_{i,k}^{(m)^{\rm H}}\big),\\
\bar v_{i,k} &\triangleq \frac{1}{M} \sum_{m=1}^{M} v_{i,k}^{(m)} = \frac{1}{M} \sum_{m=1}^{M}\big(\log _{2}(\omega_{i, k}^{(m)})\big).
\end{aligned}
\end{equation}
And similarly, we also define
\begin{equation}\label{vars_kj_def}
\begin{aligned}
\bar t_{k,j} &\triangleq \frac{1}{M} \sum_{m=1}^{M}t_{k,j}^{(m)} =\frac{1}{M} \sum_{m=1}^{M}\big(\omega_{k,j}^{(m)}|g_{k,j}^{(m)}|^{2}\big), \\
\bar \Psi_{k,j} &\triangleq \frac{1}{M} \sum_{m=1}^{M} \Psi_{k,j}^{(m)} = \frac{1}{M} \sum_{m=1}^{M} \big(t_{k,j}^{(m)} {\mathbf h}_{j}^{(m)} {\mathbf h}_{j}^{(m)H}\big), \\
\bar {\mathbf{f}}_{k,j} &\triangleq \frac{1}{M} \sum_{m=1}^{M} \mathbf{f}_{k,j}^{(m)} =\frac{1}{M} \sum_{m=1}^{M} \big(\omega_{k,j}^{(m)} \mathbf{h}_{j}^{(m)} g_{k,j}^{(m)^{\rm H}}\big),\\
\bar v_{k,j} &\triangleq \frac{1}{M} \sum_{m=1}^{M} v_{k,j}^{(m)} = \frac{1}{M} \sum_{m=1}^{M}\big(\log _{2}(\omega_{k,j}^{(m)})\big).
\end{aligned}
\end{equation}
Thus, with the optimal precoder obtained from the $[n-1]$-th iteration, ${\bf G}^{[n]}$ and ${\bf \Omega}^{[n]}$ are designed based on \eqref{g_def} and \eqref{W_def} in the outer loop. In the inner loop, $\bar{\bf x}^{[n]}$ and ${\bf P}^{[n]}$ are optimized by solving problem \eqref{P1}. Notably, under fixed ${\bf G}^{[n]}$ and ${\bf \Omega}^{[n]}$, solving problem \eqref{P1} is still challenging. The difficulty lies in the non-convex constraint \eqref{P2.3.b}. To address this, we introduce SCA method and linearly approximate $\xi_{k,j}$ by using its first-order Taylor expansion
\begin{equation}\label{xi_approx}
\begin{aligned}
{\xi}_{k,j} &\approx \omega_{k,j}|g_{k,j}|^{2}\big(\sum_{i\ne j}(2\Re{\{{\bf p}_i^{[n-1]{\rm H}}{\bf h}_j {\bf h}_j^{\rm H} {\bf p}_i\}}-|{\bf h}_j^{\rm H}{\bf p}_i^{[n-1]}|^2)\big)\\
&+\omega_{k,j}|g_{k,j}|^{2}\sigma_n^2-2\omega_{k,j}\Re{\{g_{k,j}{\bf h}_j^{\rm H} {\bf p}_k\}}+\omega_{k,j}-\log_2 (\omega_{k,j}).
\end{aligned}
\end{equation}
Therefore, at the $[n]$-th iteration, by introducing a vector $\bar{\boldsymbol\eta}=\{{\bar \eta_{k,j}}|  k\in\mathcal{K}, j\in\mathcal{K}\backslash\{k\}\}$, the inner-loop problem for optimizing $\bf P$ and $\bar{\bf x}$ is formulated as
\begin{subequations}\label{P2.5}
\begin{align}
\min_{\bf{\bar x}, \bf{P}, \bar{\boldsymbol\alpha}} & \sum\nolimits_{k \in \mathcal{K}} u_{k}(\bar{\xi}_{{\mathrm p},k}+\bar X_{k}) \tag{\ref{P2.5}}\\
\text { s.t. } & \bar{\xi}_{{\mathrm p},k}-\bar\eta_{k,j}\leq -R_{{\mathrm s},k}^{\rm th}, \forall k \in \mathcal{K}, \\
\begin{split} 
&\sum\nolimits_{i\ne k}\big( 2\Re{\{{\bf p}_i^{[n-1]{\text H}}{\bar \Psi}_{k,j}{\bf p}_i\}}-{\bf p}_i^{[n-1]{\text H}}{\bar\Psi}_{k,j}{\bf p}_i^{[n-1]} \big)  \\
& +\bar t_{k,j} \sigma_n^2 -2\Re{ \{{{\bar{\bf f}}_{k,j}^{\text H} {\bf p}_k }\}}+\bar u_{k,j}-\bar v_{k,j} \geq {\bar \eta}_{k,j} ,
\end{split}\\
& -\sum_{j \in \mathcal{K}} \bar X_{j} +\bar{\xi}_{{\mathrm c},k} \leq 1, \forall k \in \mathcal{K}, \\
& \nonumber\eqref{P0.e},\eqref{P2.3.e},
\end{align}
\end{subequations}
which is a convex problem and can be readily solved using toolboxes like CVX. The proposed algorithm for imperfect CSIT setup is summarized in Algorithm 2.

\begin{algorithm}[!t]
\caption{Joint WMMSE and SCA based AO algorithm for imperfect CSIT}
\label{Algorithm 2}
\begin{algorithmic}[1]
    \Require
    the tolerance $\epsilon_1$, $\epsilon_2$, the secrecy threshold $R_{{\mathrm s},k}^{\rm th}$, the power constraint $P_{max}$, the number of samples $M$.
    \Ensure
     the optimal secure precoder ${\mathbf P}^{\rm opt}$ and common rate vector ${\mathbf c}^{\rm opt}$. 
    \State Set outer-loop threshold $\epsilon_1$, outer-loop iteration index $n:=0$;\\
     Initialize ${\mathbf P}^{[0]}$ and the obtained WASR value $\text{WASR}^{[0]}$;
    \Repeat:
    \State Update ${\bf G}^{[n]}, {\bf \Omega}^{[n]}$ using ${\mathbf P}^{[n-1]}$ based on \eqref{g_def} and \eqref{W_def} in the outer loop;
    \State Set inner-loop threshold $\epsilon_2$, inner-loop iteration index $m:=0$;
    \Repeat:
    \State update ${\mathbf P}^{[m]}, {\bf {\bar c}}^{[m]}$ using ${\bf G}^{[n]}, {\bf \Omega}^{[n]}, {\mathbf P}^{[n-1]}$ by solving problem $\rm P2.5$ in the inner loop.
    \Until \text{the inner-loop objective ${\rm WMMSE}^{[m]}$ satisfying:}
    \begin{equation}
    \nonumber
    | {\rm WMMSE}^{[m]}-{\rm WMMSE}^{[m-1]} |\le \epsilon_2;
    \end{equation} 
    \State \Return inner-loop solution ${\mathbf P}^{*}={\mathbf P}^{[m]}$, ${\mathbf c}^{*}={\mathbf c}^{[m]}$;
    \State Update: ${\mathbf P}^{[n]}\leftarrow{\mathbf P}^{*}$, ${\mathbf c}^{[n]}\leftarrow{\mathbf c}^{*}$, $n\leftarrow n+1$;
    \Until \textit{Stopping criterion satisfying:}
    \begin{equation}
    \nonumber
    | {\text{WASR}}^{[n+1]}-{\text{WASR}}^{[n]} |\le \epsilon_1;
    \end{equation}
    \State \Return ${\mathbf P}^{\rm opt}={\mathbf P}^{[n+1]}$, ${\mathbf c}^{\rm opt}={\mathbf c}^{[n+1]}$.
\end{algorithmic}
\end{algorithm}

\emph{Initialization:}
For the perfect CSIT, the precoder ${\bf{P}}^{[0]}$ is initialized by following \cite{Hamdi2016RSWMMSE}.
\footnote{Numerical results in \cite{Hamdi2016RSWMMSE} have shown that such initialization guarantees a good suboptimal solution.} 
The initialized precoder for common messages is given by ${\bf p}_{\mathrm c}^{[0]}=p_{\mathrm c} {\bf u}_{\mathrm c}$, where  $p_{\mathrm c}=\kappa P_{max}$ and $0\leq\kappa\leq 1$. ${\bf u}_{\mathrm c}$ is the largest left singular vector of the channel matrix ${\bf H}=[ {\bf h}_1, ..., {\bf h}_K]$. 
The initialized precoders for private messages ${\bf p}_k^{[0]}$ are given by ${\bf p}_k^{[0]}=p_k\frac{{\bf h}_k}{\lVert{\bf h}_k\rVert}$, where $p_k=\frac{(1-\kappa) P_{max}}{K}$. 
The common rate vector ${\bf{c}}^{[0]}$ is initialized by assuming uniformly allocated common rate $R_{{\mathrm c},k}({\bf{P}}^{[0]})$ to all users. 
$\beta_{{\mathrm c},k}^{[0]}$, $\beta_{{\mathrm p},k}^{[0]}$ and $\alpha_{k,j}^{[0]}$ are initialized by replacing the inequalities in \eqref{11b}, \eqref{11c} and \eqref{8d} with equalities, respectively.  
The initialization for the precoder ${\bf P}^{[0]}$ under the imperfect CSIT  follows a similar way except for replacing the real channel matrix $\bf H$ by the estimated channel matrix $\widehat{\bf H}$ and no additional introduced parameters are needed.

\emph{Convergence analysis:}
We first prove the convergence of Algorithm 2. As the AO algorithm goes on, the optimal solution $({\bf P}^{[n-1]}$, ${\bar {\bf x}}^{[n-1]}$, ${\bf \Omega}^{[n-1]}$, ${\bf G}^{[n-1]})$ obtained at the $[n-1]$-th iteration also serves as a feasible solution at the $[n]$-th iteration. Correspondingly, for the problem \eqref{P1}, the derived objective value at the $[n]$-th iteration is no larger than that of $[n-1]$-th. Since the power constraint holds, the objective function in \eqref{P1} is bounded and monotonically decreasing as iterations carry on. Therefore, the joint WMMSE and SCA based AO algorithm is guaranteed to converge.

Based on \eqref{Rate-WMMSE}, it is observed that at the $[n]$-th iteration, problem \eqref{P2.5} is a convex approximation of the sampled optimization problem \eqref{P2.2_SAA} at the solution ${\bf P}^{[n-1]}$. 
Therefore, following the analysis in \cite{Hamdi2016RSWMMSE}, the proposed joint WMMSE and SCA based algorithm is also a special SCA method. 
Considering the fact that \cite[Assumption 1]{razaviyayn2014successive} is satisfied and the solution sequence obtained by Algorithm 2 in the AO iterations lies in a compact feasible set, the solutions of Algorithm 2 converge to the set of KKT points of the problem \eqref{P2.2_SAA} according to \cite[Theorem 1]{razaviyayn2014successive}. Besides, since ${\xi}_{k,j}$ is approximated by its first-order upper bound, according to \cite[Theorem 1]{razaviyayn2013unified} and \cite[Corollary 1]{razaviyayn2013unified}, and considering the fact that the AO procedure is also an instance of SCA method, every limit point  generated by solving problem \eqref{P2.5} is a stationary point of problem \eqref{P1} and therefore a stationary point of problem \eqref{P2.2_SAA} due to the fact that problem \eqref{P2.2_SAA} is equivalent to \eqref{P1}.
As $M \to \infty$, problem \eqref{P2.1_ASRform} is equvilent to \eqref{P2.2_SAA}, therefore, the convergent point of the problem \eqref{P2.2_SAA} is also a stationary point of problem \eqref{P2.1_ASRform}. 

\section{Numerical Results}
In this section, we evaluate the WSR performance of the considered MISO BC secure communication model. 
The most typically compared multiple access baseline strategies are OMA, multicasting, NOMA, and SDMA (SDMA is referred to as MULP in our work). For OMA, only one user is served per channel use, therefore, no internal eavesdropper exists. For multicasting, user data is broadcast to all users in the user group, thus no security can be obtained among these users. 
For NOMA, since one whole user message is mapped into the common stream, this message is entirely decoded by other users and NOMA cannot guarantee all users' secrecy rate constraints. As a result, although NOMA is a baseline, only MULP can be considered in the numerical evaluations.  

Assume the channel noise variance $\sigma_n^2=1$. Therefore, the transmit SNR calculated by ${\rm SNR} \buildrel \Delta \over = \frac{P_{max}}{\sigma_n^2}$ is equal to the value of $P_{max}$. 
Set the transmit SNR as 20dB, unless otherwise stated. 
We also assume all users have the same secrecy rate constraint, i.e., $R_{{\rm s},k}^{\rm th}=R_{{\rm s}}^{\rm th}, \forall k\in\cal{K}$. 
For the imperfect CSIT case, the size of channel error samples is set to $M=1000$ throughout simulations. 
We set the error covariance of each user $\sigma_{{\rm e},k}^2=\sigma_{\rm e}^2= \gamma_{\rm e} P_{max}^{-\delta}, \forall k\in\cal{K}$, where $\gamma_{\rm e}$ represents the CSIT qualities, and the scaling factor $\delta=0.6$. 
To construct the channel realization samples $\mathbb{H}^{(M)}$, we set the $m$th conditional channel realization as ${\bf H}^{(m)}=\sqrt {1-\sigma_{\rm e}^2}{\bf\widehat H}+\sqrt {\sigma_{\rm e}^2}{\bf\widetilde H}^{(m)}$ \cite{Hamdi2016RSWMMSE}. 


\begin{figure}[h]
\centerline{\includegraphics[scale=0.7]{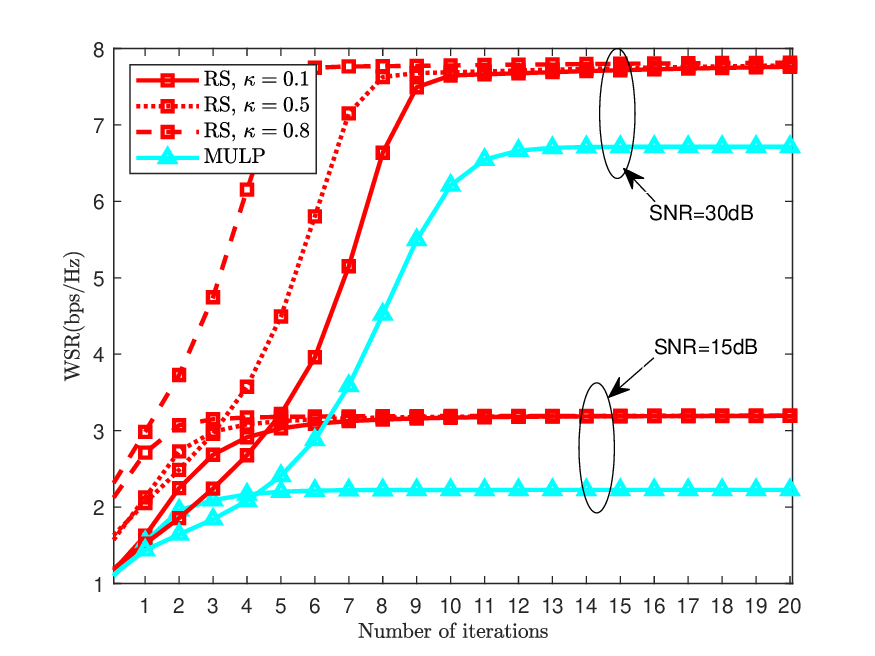}}
\caption{The WSR convergence using Algorithm 2 for 3 different values of initialization parameter $\kappa$ under a specific channel, $N_{\rm t}=2, \gamma=1, \theta=\frac{2\pi}{9}$, $R_{\rm s}^{\rm th}=0.5$. }
\label{convergence_s}
\end{figure}
We first evaluate the convergence of the proposed algorithm 2. Fig. \ref{convergence_s} illustrates the WSR convergence of Algorithm 2 for both RSMA-based secure precoding and MULP schemes under a specific channel, where the initialization parameter $\kappa$ is set to 0.1, 0.5, 0.8, SNR=15dB, 30dB, and $N_{\rm t}=2, \gamma=1, \theta=\frac{2\pi}{9}$, $R_{\rm s}^{\rm th}=0.5$. 
Compared with MULP, RS converges to a better WSR value regardless of the initialization parameter. 
As iterations carry on, the algorithm converges to a limit point for both schemes no matter what value the initialization parameter $\kappa $ is. 
However, for RS, the convergence speed may be different under different  $\kappa$.

\subsection{Simulation results for specific channels}
To investigate the influence of user angles and channel strength on WSR performance, the specific-channel scenario \cite{mao2018rsma} is first analyzed. 
When $N_{\rm t}=2$, the user channels are realized as
\begin{equation}
{\bf{h}}_1=[1, 1]^{\rm H},\quad{\bf{h}}_2=\gamma\times[1, e^{j\theta}]^{\rm H},
\end{equation} 
where $\gamma$, $\theta$ represents the relative channel strength and angle difference between user-2 and user-1.  
Assuming $\gamma_{\rm e}=\gamma$ \cite{mao2018rsma},
we focus on the 2-user scenario in this subsection and investigate the WSR performance when $\theta \in \{ \frac{\pi}{9}, \frac{2\pi}{9}, \frac{3\pi}{9}, \frac{4\pi}{9} \}$. The user weight vector is set as ${\bf u}=[0.5, 0.5]$.
     
\begin{figure}[t!]
\begin{minipage}[t]{0.48\linewidth}
    \includegraphics[scale=0.4]{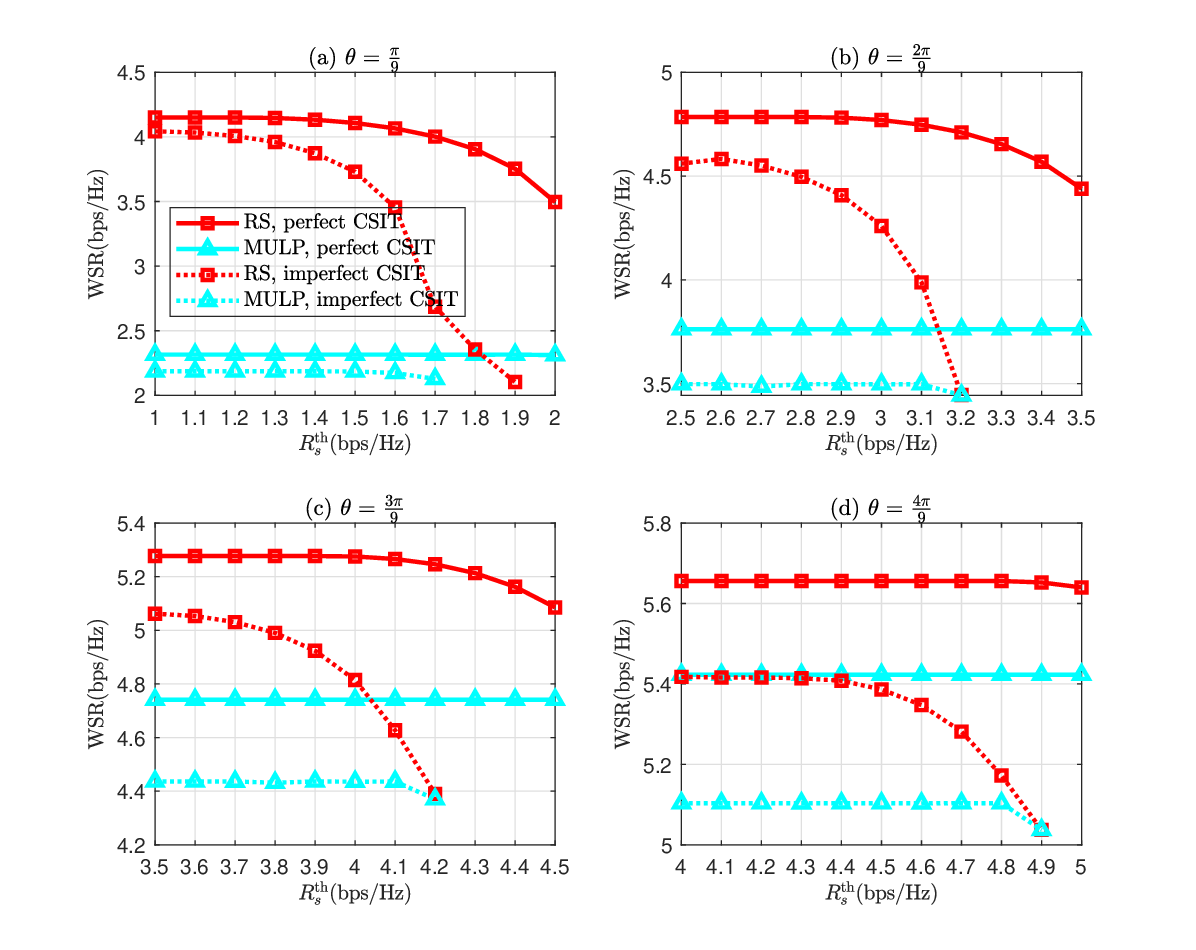}
    \caption{The WSR vs $R_{{\rm s}}^{\rm th}$ for secure RS and MULP schemes under perfect and imperfect CSIT, SNR=20dB, $N_{\rm t}=2, \gamma=1$.}
    \label{(2,2)_specific_WSR}
\end{minipage}%
    \hfill%
\begin{minipage}[t]{0.48\linewidth}
    \includegraphics[scale=0.42]{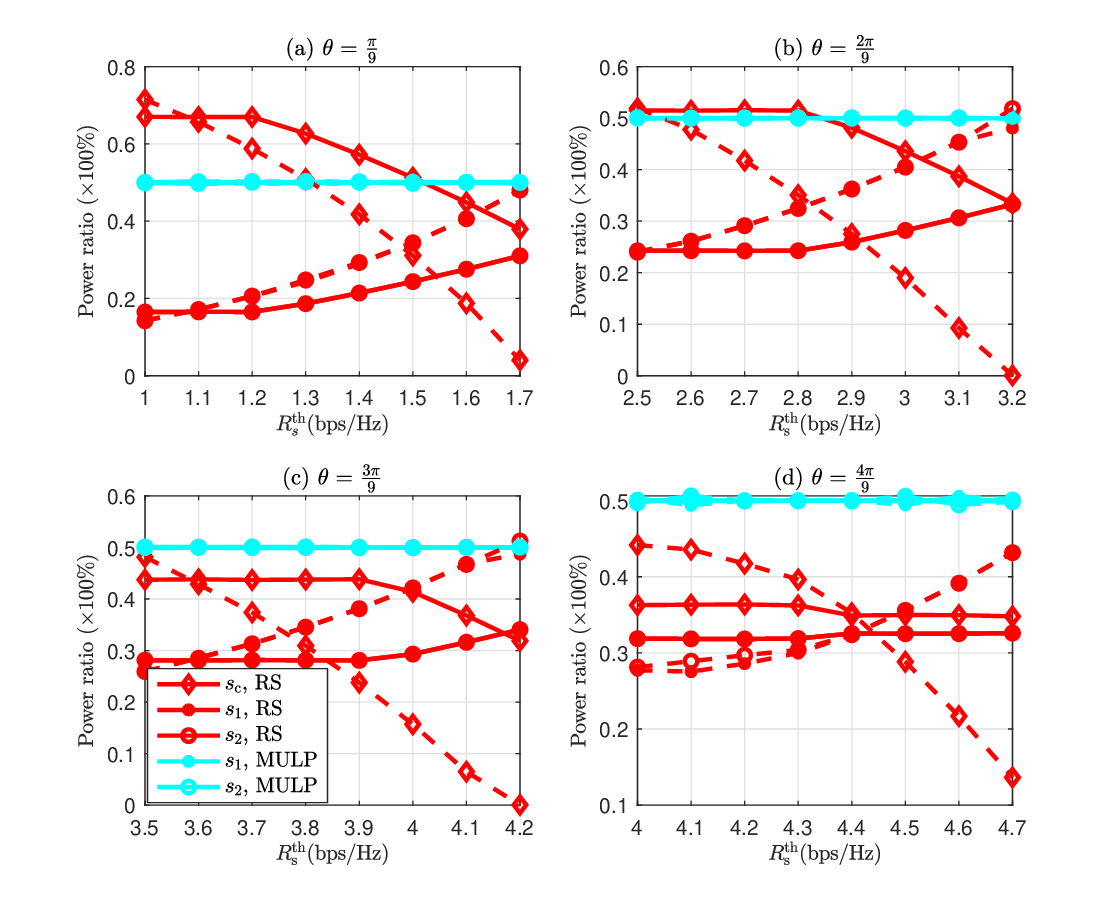}
	\caption{The power allocation to data streams vs $R_{s}^{\rm th}$ under perfect and imperfect CSIT, SNR=20dB, $N_t=2, \gamma=1$.}
	\label{(2,2)_specific_PA}
\end{minipage} 
\end{figure}

Fig. \ref{(2,2)_specific_WSR} illustrates the WSR performance of the secure RS and MULP schemes with respect to secrecy rate threshold $R_{{\rm s}}^{\rm th}$, where SNR=20dB, $N_{\rm t}=2$, $\gamma=1$. The communication system is strictly underloaded and both user channels have the same channel strength. 
For both perfect and imperfect CSIT, the WSR of the proposed RSMA-based secure precoding always outperforms that of MULP as $\theta$ ranges from $\frac{\pi}{9}$ to $\frac{4\pi}{9}$. 
This is because RS is more flexible in optimizing precoders by bridging common rate and private rate. 
Specifically, the WSR of secure RS first remains unchanged then decreases, because when $R_{{\rm s}}^{\rm th}$ is relatively small, RS is capable of ensuring security by optimizing private streams without suffering from WSR loss. 
As $R_{{\rm s}}^{\rm th}$ increases, more power should be allocated to private streams to enhance security and meanwhile a WSR loss is introduced. 
This phenomenon is especially obvious under imperfect CSIT, even though a higher performance gain can be achieved in this case when $R_{\rm s}^{\rm th}$ is small. 
Besides, the RS scheme can satisfy higher or at least equal security requirements, although it is more sensitive to $R_{\rm s}^{\rm th}$ changes due to the power allocated to common streams should adapt to the $R_{\rm s}^{\rm th}$ changes. 
Comparatively, the WSR of MULP still remains at a relatively low level compared with RS as no common streams are scheduled. 

Fig. \ref{(2,2)_specific_PA} illustrates the distribution of allocated power to the data streams $s_{\rm c}$, $s_{1}$ and $s_{2}$ for RSMA-based secure precoding and MULP scheme under both perfect CSIT and imperfect CSIT. 
The power ratio curves of private streams $s_{1}$ and $s_{2}$ are overlapped for both MULP and RS, since $\gamma=1$ indicates both users have the same channel strength. 
When perfect CSIT is available, as $R_{\rm s}^{\rm th}$ increases, the power allocated to $s_{\rm c}$ of RS scheme first remains constant then decreases under all $\theta$ values. 
Whereas the power ratio of $s_{\rm c}$ under imperfect CSIT condition keeps decreasing and eventually approaches 0, this is when the WSR of RS and MULP coincide with each other. 
This indicates a higher $R_{\rm s}^{\rm th}$ requires more power allocated to private streams, which is in line with results in Fig. \ref{(2,2)_specific_WSR}. 
Correspondingly, the power allocated to private streams under imperfect CSIT is higher than under perfect CSIT. 
Because a poor channel condition tends to allocate more power to private streams to ensure security. 
As there is no common stream in MULP, the power ratio for both private streams stay constant at 0.5. 


\begin{figure}[htbp!]
\begin{minipage}[t]{0.48\linewidth}
	\includegraphics[scale=0.415]{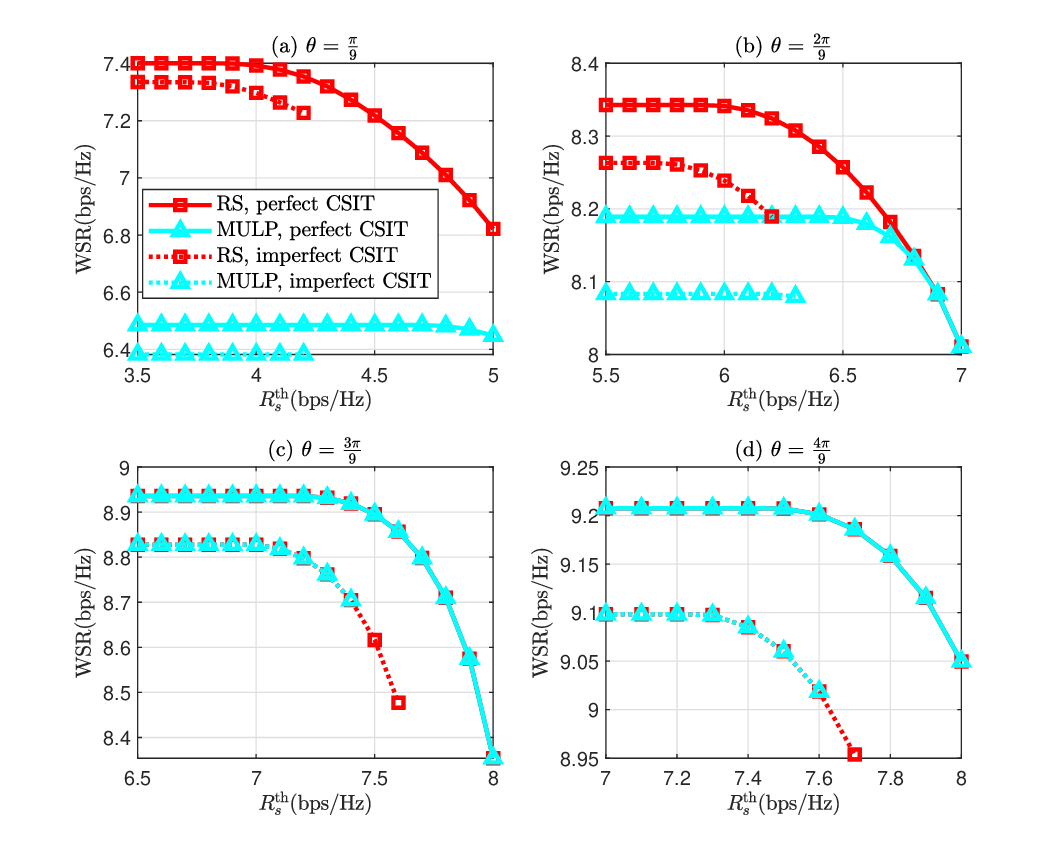}
 	\caption{The WSR vs $R_{{\rm s}}^{\rm th}$ for secure RS and MULP schemes, SNR=30dB, $N_{\rm t}=4, \gamma=0.3$.}
 	\label{(4,2)_specific_b0.3_WSR}
\end{minipage}%
    \hfill%
\begin{minipage}[t]{0.48\linewidth}
	\includegraphics[scale=0.42]{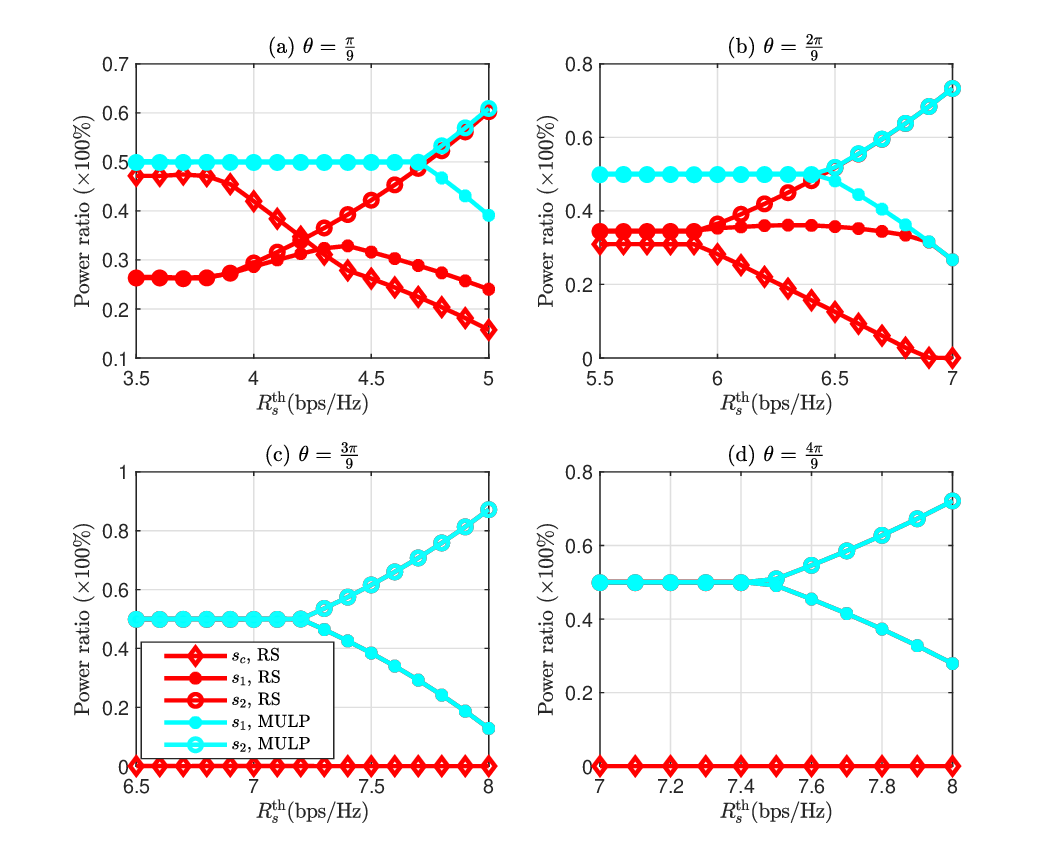}
	\caption{The power allocation to data streams vs $R_{{\rm s}}^{\rm th}$, SNR=30dB, $N_{\rm t}=4, \gamma=0.3$.}
	\label{(4,2)_specific_b0.3_PA}
\end{minipage} 
\end{figure}



Fig. \ref{(4,2)_specific_b0.3_WSR} and Fig. \ref{(4,2)_specific_b0.3_PA} illustrate the WSR and the power distribution of data streams with respect to $R_{{\rm s}}^{\rm th}$ for RSMA-based secure precoding and MULP schemes, where SNR=30dB, $N_{\rm t}=4$, $\gamma=0.3$. User-2 has 10 dB channel strength difference compared with user-1. 
For clarity, we omit the power allocation for imperfect CSIT.
From Fig. \ref{(4,2)_specific_b0.3_WSR}, RS exhibits a performance advantage over MULP when user channels are relatively aligned to each other, i.e., $\theta=\frac{\pi}{9}$. 
Besides, compared with the perfect CSIT case, RS achieves a higher performance advantage under imperfect CSIT when $\theta=\frac{2\pi}{9}$, this is because the RS design is more robust to channel errors than the MULP design. 
Whereas when $\theta=\frac{3\pi}{9}$ and $\theta=\frac{4\pi}{9}$, these two schemes achieve the same WSR under both CSIT conditions. 
This shows as $\theta$ increases, channel orthogonality dominates in securing data streams. 
Regarding the power allocation, from Fig. \ref{(4,2)_specific_b0.3_PA}, the power allocated to $s_{\rm c}$ first remains constant then decreases as $R_{{\rm s}}^{\rm th}$ increases when $\theta=\frac{\pi}{9}$ and $\theta=\frac{2\pi}{9}$. 
This also explains why RS outperforms MULP in WSR performance in these cases.
Meanwhile, with fixed $\theta$, a higher $R_{{\rm s}}^{\rm th}$ tends to allocate more power to the private stream of the user with poorer channel strength. 
When user channels are approximately orthogonal to each other, both schemes obtain the same power allocation results, hence their corresponding WSR curves coincide.

\begin{figure}[h]
\centerline{\includegraphics[scale=0.7]{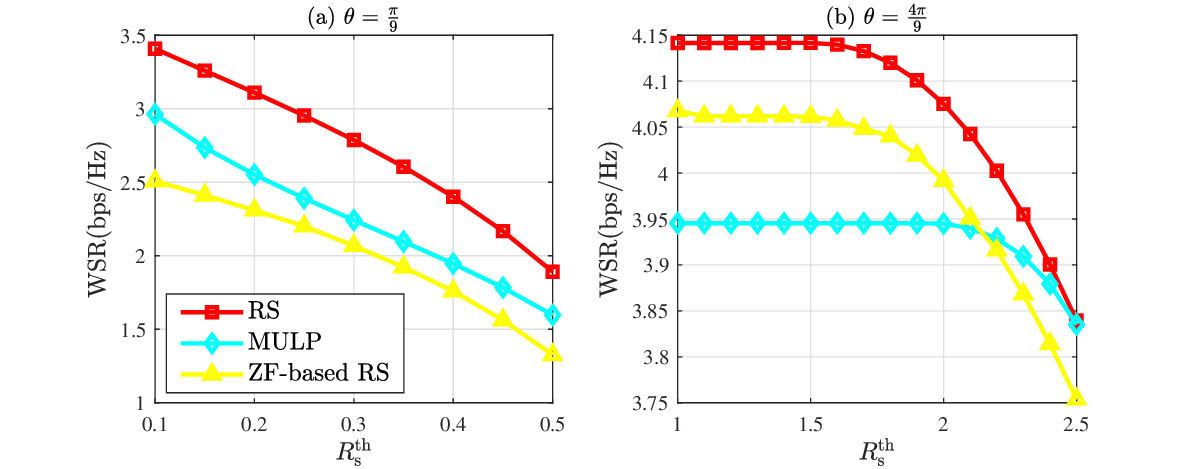}}
\caption{The WSR versus $R_{\rm s}^{\rm th}$ for 2-user secure communications under specific channels, SNR=20dB, $N_{\rm t}=2$, $\gamma=0.3$, (a) $\theta=\frac{\pi}{9}$, (b) $\theta=\frac{4\pi}{9}$.}
\label{ZF_specific}
\end{figure}
To gain more insights, by extending the beamforming strategy in \cite{bruno2020ZF}, a new baseline algorithm is presented in this simulation, denoted by the ZF-based RS.
For the ZF-based RS, the precoding directions for the private streams are fixed and designed by ZF, the precoding direction for the common stream and the power allocation among all streams are jointly optimized. 
The obtained WSR performance is shown in Fig. \ref{ZF_specific}. 
We observe that the proposed RSMA-based secure design outperforms the baseline schemes in both subfigures. 
Besides, when user channels are relatively aligned, i.e., $\theta=\frac{\pi}{9}$, the ZF-based RS performs worse than MULP for all secrecy thresholds, since the former has to sacrifice WSR performance for fully eliminating the MUI. 
However, when $\theta=\frac{4\pi}{9}$, the ZF-based RS first outperforms MULP then performs worse than MULP. Because when user channels are relatively orthogonal, the ZF-based RS is capable of ensuring MUI elimination without introducing extra WSR loss. However, with $R_{\rm s}^{\rm th}$ increasing, eliminating all MUI by the fixed precoding direction is no longer a better choice than MULP. This again proved that, by managing MUI, RS is capable of achieving better WSR while ensuring security.

\subsection{Simulation results for random channels}
In this subsection, we investigate the WSR and WESR performance of the secure RS and MULP schemes under random channels for imperfect CSIT cases. 
Each element of the user channel follows the independent and identical zero-mean complex Gaussian distribution with unit variance, i.e., $\mathcal{CN}(0, 1)$.
For random channels, 100 i.i.d. channel realizations are generated and the ones which are feasible for a specific range of $R_{\rm s}^{\rm th}$ are kept in each simulation setting. 

\begin{figure}[htbp]
\centerline{\includegraphics[scale=0.7]{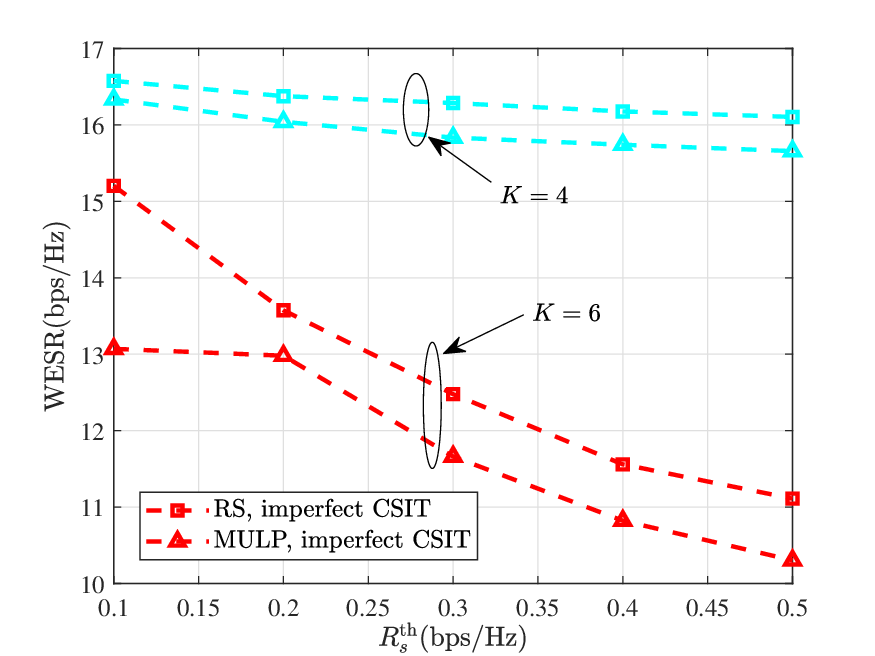}}
\caption{The WESR versus $R_{s}^{\rm th}$ over 100 random channels for secure RS and MULP designs, SNR=20dB, $N_t=4$, $K=4,6$.}
\label{WESR_Rth_Kusers}
\end{figure}
Fig. \ref{WESR_Rth_Kusers} illustrates the WESR performance with respect to $R_{\rm s}^{\rm th}$ under imperfect CSIT for both underloaded case, i.e., $N_t=K=4$, and overloaded case, i.e., $N_t=4$, $K=6$. The transmit SNR is 20dB and the weight of each user is set to 1, $\delta=0.6$.  
We observe that, for both cases, the RSMA-based secure beamforming achieves a WESR performance improvement compared with secure MULP. This comes from the fact that the existence of the common stream in RS enables a flexible MUI management.
Specifically, for both cases, the WESR performance decreases as $R_{\rm s}^{\rm th}$ increases. This means, adding secrecy constraints into both schemes would introduce a WESR performance loss compared with no secrecy requirement case.
Besides, compared with the underloaded scenario, the advantage of secure RS design over the MULP design in overloaded scenario is higher. This indicates that the proposed secure RS design is more capable of serving multiple users at a relatively high WESR performance while guaranteeing secure communication of confidential messages.

\begin{figure}[htbp]
\centerline{\includegraphics[scale=0.7]{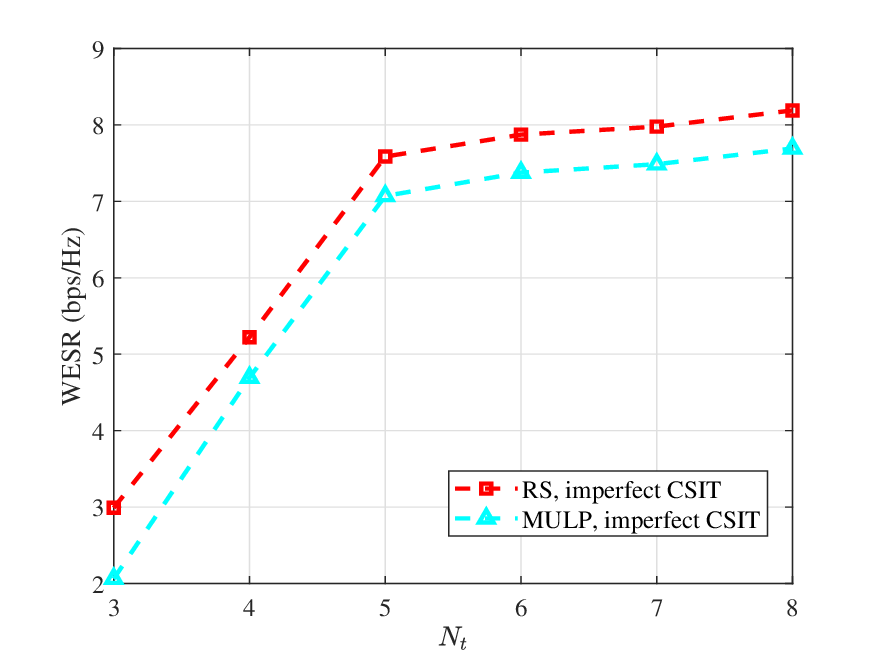}}
\caption{The WESR of RS and MULP designs versus the number of transmit antennas $N_t$ under imperfect CSIT,  $\delta=0.6$, SNR=30dB, $R_{\rm s}^{\rm th}=1$ bps/Hz.}
\label{WESR_Nts}
\end{figure}
Fig. \ref{WESR_Nts} illustrates the WESR performance of the RS and MULP schemes under imperfect CSIT setups for different number of transmit antennas, where SNR=30dB, $R_{\rm s}^{\rm th}=1$ bps/Hz and $\delta=0.6$. From Fig. \ref{WESR_Nts}, the WESR performance of both cases grows as the number of transmit antennas $N_t$ increases. Besides, the RS design still performs better than MULP designs in terms of WESR and the performance gap first narrows then becomes stable as $N_t$ increases. Specifically, for the overloaded case when $N_t=3$, the performance benefit of the RS design over the MULP design is large. However, as $N_t$ increases to 4, the advantage of RS decreases. Then as $N_t$ grows from 5 to 8, the performance gap between the RS design and the MULP design is stable. This simulation result coincides with the result in Fig. \ref{WESR_Rth_Kusers}, which shows the RS design reaps higher advantages over the MULP design in overloaded scenarios.

\begin{figure}[htbp]
\centerline{\includegraphics[scale=0.7]{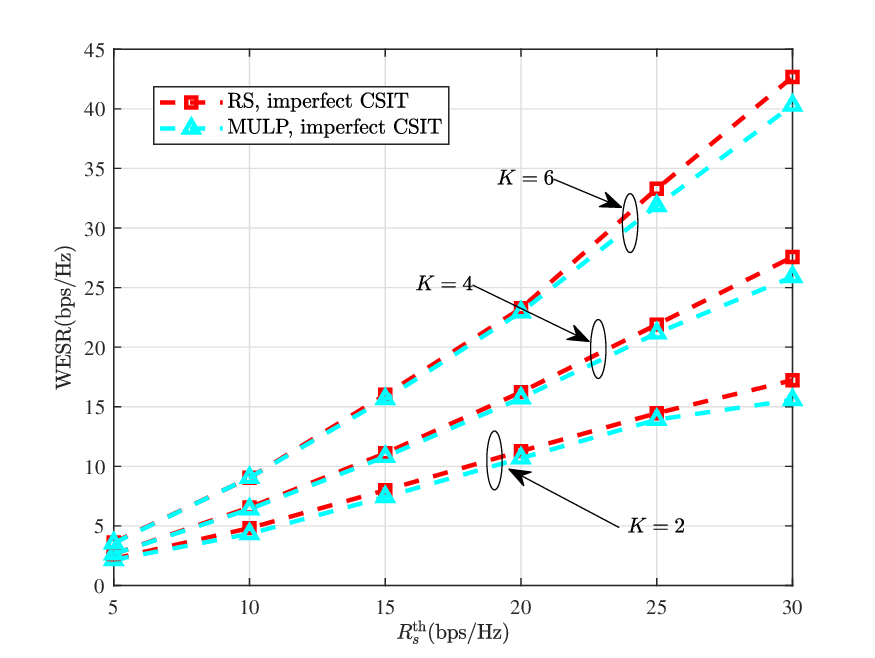}}
\caption{The WESR of RS and MULP designs versus SNR with  $\delta=0.6$, $R_{\rm s}^{\rm th}=0.5$bps/Hz and $N_t=K=2,4,6$.}
\label{WESR_vs_SNRdBs}
\end{figure}
Fig. \ref{WESR_vs_SNRdBs} illustrates the WESR performance of the RS and MULP schemes under imperfect CSIT setups for $K=2,4$ and 6 users with $\delta=0.6$. The secrecy rate threshold is set as $R_{\rm s}^{\rm th}=0.5$bps/Hz. Fig. \ref{WESR_vs_SNRdBs} demonstrates that the proposed RS design performs better than the MULP design in terms of WESR and the performance gap enlarges as SNR increases.

\section{Conclusion}
In this paper, the weighted sum-rate maximization design in RSMA-based MISO BC with confidential messages was investigated. 
Considering user security and power constraint, we formulated an instantaneous WSR maximization problem under perfect CSIT and a WESR maximization problem under imperfect CSIT. 
To handle the non-convex constraints and optimize the secure precoder, an SCA-based algorithm and a joint WMMSE and SCA based AO algorithm were proposed to solve the respective problem. 
Numerical results illustrated that, benefiting from the capability of partially decoding MUI and partially treating MUI as noise, the WSR performance of the RSMA-based secure design always outperforms the baseline schemes while guaranteeing user security.
Furthermore, compared with baseline schemes, RS is more robust to channel errors and achieves higher WSR performance gain under imperfect CSIT. 
Finally, the performance advantage achieved by the proposed RS design over the baseline algorithm enlarges in high SNR regime and overloaded cases.

\appendices
\section{Equivalence proof}
First, for the sake of simplicity, the SAA-based averaged values are replaced by their corresponding instantaneous values in the equivalence proof, since it does not influence the equivalence between problem \eqref{P1} and \eqref{P2.2_SAA}. 
For problem \eqref{P1}, the Lagrange function is constructed as 
\begin{equation}
\begin{aligned}
F({\bf P}, {\bf x}, {\bf \Omega}, {\bf G}, {\bf \Lambda})&= \sum_{k \in \mathcal{K}} u_{k} ({\xi}_{k}+ X_{k} )+\sum_{k\in\cal{K}}{\sum_{j\in \mathcal{K} \backslash\{k\}}\phi_{k,j}} \\
&+\sum_{k\in\mathcal{K}}\psi_{{\rm c},k}+\lambda_0[\operatorname{tr}(\mathbf{P} \mathbf{P}^{\rm H})-P_{{\rm t}}],
\end{aligned}
\end{equation}
where $\phi_{k,j}=\lambda_{k,j}(\xi_{{\rm p},k}-\xi_{k,j}+R_{{\mathrm s},k}^{\rm th})$, $\psi_{{\rm c},k}=\lambda_{{\rm c},k}(-\sum_{i \in \mathcal{K}}  X_{i} +{\xi}_{{\rm c},k}^{(M)}-1)$, ${\bf \Lambda}=\{\lambda_{k,j},\lambda_{{\rm c},k},\lambda_0\},\forall k\in{\mathcal{K}}$.
Assume the optimal solution for \eqref{P1} is $({\bf P}^*, {\bf x}^*, {\bf \Omega}^*, {\bf G}^*, {\bf \Lambda}^*)$. From the KKT conditions \cite{Christensen2008WMMSE}, the optimal solution ${\bf G}^*$ and ${\bf \Omega}^*$ must satisfy:
\begin{subequations}\label{lag_condition}
\begin{align}
{\left. {\frac{{\partial F}}{{\partial G}}} \right|_{G= {G^*}}} &= 0, \label{lag_1}\\
{\left. {\frac{{\partial F}}{{\partial \Omega }}} \right|_{\Omega ={\Omega ^*}}}& = 0. \label{lag_2}
\end{align}
\end{subequations}
Since $\xi_{{\rm c},k}$, $\xi_{{\rm p},k}$ and $\xi_{k,j}$ are independent from each other, 
\begin{subequations}\label{g_derive}
\begin{align}
{\frac{{\partial F}}{{\partial g_{{\rm c},k}}}}&=\lambda_{{\rm c},k}{\frac{{\partial \xi_{{\rm c},k}}}{{\partial g_{{\rm c},k}}}}, \\
{\frac{{\partial F}}{{\partial g_{{\rm p},k}}}}&=(\mu_{k}+\sum_{j\in\mathcal{K}\backslash\{k\}}\lambda_{k,j}){\frac{{\partial \xi_{{\rm p},k}}}{{\partial g_{{\rm p},k}}}}, \\
{\frac{{\partial F}}{{\partial g_{k,j}}}}&=-\lambda_{k,j}{\frac{{\partial \xi_{k,j}}}{{\partial g_{k,j}}}}.
\end{align}
\end{subequations}
Plugging \eqref{xi_c_trans}-\eqref{xi_kj_trans} into \eqref{g_derive} yields
\begin{subequations}
\begin{align}
{\frac{{\partial \xi_{{\rm c},k}}}{{\partial g_{{\rm c},k}}}}&=\omega_{{\rm c},k}T_{{\rm c},k}g_{{\rm c},k}^{\rm H}-\omega_{{\rm c},k}h_k^{\rm H}{\bf p}_{\rm c}, \\
{\frac{{\partial \xi_{{\rm p},k}}}{{\partial g_{{\rm p},k}}}}&=\omega_{{\rm p},k} T_{{\rm p},k} g_{{\rm p},k}^{\rm H}-\omega_{{\rm p},k} h_k^{\rm H} {\bf p}_k,\\
{\frac{{\partial \xi_{k,j}}}{{\partial g_{k,j}}}}&=\omega_{j,k} I_{k} g_{j,k}^{\rm H}-\omega_{j,k} h_k^{\rm H} {\bf P}_j.
\end{align}
\end{subequations}
Since $\lambda_{{\rm c},k}>0$, $\lambda_{k,j}>0$ and $\mu_k>0$, we have 
\begin{equation}
{\frac{{\partial \xi_{{\rm c},k}}}{{\partial g_{{\rm c},k}^*}}}=0, \quad {\frac{{\partial \xi_{{\rm p},k}}}{{\partial g_{{\rm p},k}^*}}}=0, \quad {\frac{{\partial \xi_{k,j}}}{{\partial g_{k,j}^*}}}=0.
\end{equation}
The corresponding optimal solutions of MMSE equalizers are then given in \eqref{g_def}.
Similarly, 
\begin{subequations}\label{w_derive}
\begin{align}
{\frac{{\partial \xi_{{\rm c},k}}}{{\partial \omega_{{\rm c},k}}}}&=|g_{{\rm c},k}|^2 T_{{\rm c},k} - 2\Re\{g_{{\rm c},k} h_k^{\rm H}{\bf p}_{\rm c}\}+1-\frac{1}{\ln 2 \omega_{{\rm c},k}},\\
{\frac{{\partial \xi_{{\rm p},k}}}{{\partial \omega_{{\rm p},k}}}}&=|g_{{\rm p},k}|^2 T_{{\rm p},k} - 2\Re\{g_{{\rm p},k} h_k^{\rm H}{\bf p}_k\}+1-\frac{1}{\ln 2 \omega_{{\rm p},k}},\\
{\frac{{\partial \xi_{k,j}}}{{\partial \omega_{k,j}}}}&=|g_{k,j}|^2 I_{j} - 2\Re\{g_{k,j} h_j^{\rm H}{\bf p}_k\}+1-\frac{1}{\ln 2 \omega_{k,j}}.
\end{align}
\end{subequations}
Plugging \eqref{lag_2} into \eqref{w_derive}, the optimal MMSE weights are given in \eqref{W_def}. The scaling factor $\frac{1}{\ln 2}$ is omitted as it has no effect on the solution \cite{Hamdi2016RSWMMSE}. 
Now we need to prove under given optimal MMSE equalizers and weights, problem \eqref{P1} can be equivalently transformed to \eqref{P2.2_SAA} \cite{Shi2011WeightedMMSE}. For given ${\bf G}^*$ and $\Omega^*$, by plugging \eqref{xi_c_trans}-\eqref{xi_kj_trans}, the corresponding relationship between the MSEs and rates are the same as \eqref{Rate-WMMSE}.
Therefore, problem \eqref{P2.2_SAA} is equivalent to \eqref{P1} in the sense that the optimal solutions for the secure precoder and common rate vector for the two problems are identical. It implies that the WSR maximization in problem \eqref{P2.2_SAA} can be accomplished by solving problem \eqref{P1} under properly designed MMSE equalizers and weights.

\bibliographystyle{IEEEtran}
\bibliography{IEEEabrv,sRS}

\begin{thebibliography}{10}
\providecommand{\url}[1]{#1}
\csname url@samestyle\endcsname
\providecommand{\newblock}{\relax}
\providecommand{\bibinfo}[2]{#2}
\providecommand{\BIBentrySTDinterwordspacing}{\spaceskip=0pt\relax}
\providecommand{\BIBentryALTinterwordstretchfactor}{4}
\providecommand{\BIBentryALTinterwordspacing}{\spaceskip=\fontdimen2\font plus
\BIBentryALTinterwordstretchfactor\fontdimen3\font minus
  \fontdimen4\font\relax}
\providecommand{\BIBforeignlanguage}[2]{{%
\expandafter\ifx\csname l@#1\endcsname\relax
\typeout{** WARNING: IEEEtran.bst: No hyphenation pattern has been}%
\typeout{** loaded for the language `#1'. Using the pattern for}%
\typeout{** the default language instead.}%
\else
\language=\csname l@#1\endcsname
\fi
#2}}
\providecommand{\BIBdecl}{\relax}
\BIBdecl

\bibitem{Xia2022secrecyRS}
H.~Xia, Y.~Mao, B.~Clerckx, X.~Zhou, S.~Han, and C.~Li, ``Weighted sum-rate
  maximization for rate-splitting multiple access based secure communication,''
  in \emph{2022 {IEEE} Wireless Communications and Networking Conference
  ({WCNC})}, May 2022, pp. 19--24.

\bibitem{Bruno2016RS}
B.~Clerckx, H.~Joudeh, C.~Hao, M.~Dai, and B.~Rassouli, ``Rate splitting for
  {MIMO} wireless networks: {A} promising {PHY}-layer strategy for {LTE}
  evolution,'' \emph{{IEEE} Commun. Mag.}, vol.~54, no.~5, pp. 98--105, May
  2016.

\bibitem{mao2018rsma}
Y.~Mao, B.~Clerckx, and V.~O. Li, ``Rate-splitting multiple access for downlink
  communication systems: bridging, generalizing, and outperforming {SDMA} and
  {NOMA},'' \emph{EURASIP J. Wireless Commun. Netw.}, vol. 2018, no.~1, pp.
  1--54, May 2018.

\bibitem{Hamdi2016robustRS}
H.~Joudeh and B.~Clerckx, ``Robust transmission in downlink multiuser {MISO}
  systems: A rate-splitting approach,'' \emph{{IEEE} Trans. Signal Process.},
  vol.~64, no.~23, pp. 6227--6242, Dec. 2016.

\bibitem{onur2021mobility}
O.~Dizdar, Y.~Mao, and B.~Clerckx, ``Rate-splitting multiple access to mitigate
  the curse of mobility in (massive) {MIMO} networks,'' \emph{{IEEE} Trans.
  Commun.}, vol.~69, no.~10, pp. 6765--6780, Oct. 2021.

\bibitem{mao2019RSSEEE}
Y.~Mao, B.~Clerckx, and V.~O.~K. Li, ``Rate-splitting for multi-antenna
  non-orthogonal unicast and multicast transmission: Spectral and energy
  efficiency analysis,'' \emph{{IEEE} Trans. Commun.}, vol.~67, no.~12, pp.
  8754--8770, Dec. 2019.

\bibitem{Chen2020userfairness}
H.~Chen, D.~Mi, Z.~Chu, P.~Xiao, Y.~Xu, and D.~He, ``{Link-level performance of
  rate-splitting based downlink multiuser MISO systems},'' in \emph{Proc.
  {IEEE} Int. Symp. Personal Indoor Mobile Radio Commun. (PIMRC)}, Sept. 2020,
  pp. 1--5.

\bibitem{Caus2018reliability}
M.~Caus, A.~Pastore, M.~Navarro, T.~Ramirez, C.~Mosquera, N.~Noels, N.~Alagha,
  and A.~I. Perez-Neira, ``{Exploratory analysis of superposition coding and
  rate splitting for multibeam satellite systems},'' in \emph{Proc. {IEEE} Int.
  Symp. Wireless Commun. Syst. (ISWCS)}, Oct. 2018, pp. 1--5.

\bibitem{xie2021PLAsurvey}
N.~Xie, Z.~Li, and H.~Tan, ``A survey of physical-layer authentication in
  wireless communications,'' \emph{{IEEE} Commun. Surveys Tuts.}, vol.~23,
  no.~1, pp. 282--310, 1st Quart. 2021.

\bibitem{Yu2020IRSsecurity}
X.~Yu, D.~Xu, Y.~Sun, D.~W.~K. Ng, and R.~Schober, ``Robust and secure wireless
  communications via intelligent reflecting surfaces,'' \emph{IEEE Journal on
  Selected Areas in Communications}, vol.~38, no.~11, pp. 2637--2652, Nov.
  2020.

\bibitem{shannon1949secrecy}
C.~E. {Shannon}, ``Communication theory of secrecy systems,'' \emph{{Bell}
  Syst. Tech. J.}, vol.~28, no.~4, pp. 656--715, Oct. 1949.

\bibitem{wyner1975wiretap}
A.~D. {Wyner}, ``The wire-tap channel,'' \emph{{Bell} Syst. Tech. J.}, vol.~54,
  no.~8, pp. 1355--1387, Oct. 1975.

\bibitem{Csiszar1978BC}
I.~Csiszar and J.~Korner, ``Broadcast channels with confidential messages,''
  \emph{{IEEE} Trans. Inf. Theory}, vol.~24, no.~3, pp. 339--348, 1978.

\bibitem{Parada2005P2P}
P.~Parada and R.~Blahut, ``{Secrecy capacity of SIMO and slow fading
  channels},'' in \emph{Proc. {IEEE} Int. Symp. Inf. Theory (ISIT) Workshop},
  2005, pp. 2152--2155.

\bibitem{Geraci2012RCI}
G.~Geraci, M.~Egan, J.~Yuan, A.~Razi, and I.~B. Collings, ``{Secrecy sum-rates
  for multi-user MIMO regularized channel inversion precoding},'' \emph{{IEEE}
  Trans. Commun.}, vol.~60, no.~11, pp. 3472--3482, Nov. 2012.

\bibitem{hao2020robustsecureRS}
H.~Fu, S.~Feng, W.~Tang, and D.~W.~K. Ng, ``Robust secure beamforming design
  for two-user downlink {MISO} rate-splitting systems,'' \emph{{IEEE} Trans.
  Wireless Commun.}, vol.~19, no.~12, pp. 8351--8365, Dec. 2020.

\bibitem{Ping2020cooperativeRS}
P.~Li, M.~Chen, Y.~Mao, Z.~Yang, B.~Clerckx, and M.~Shikh-Bahaei, ``Cooperative
  rate-splitting for secrecy sum-rate enhancement in multi-antenna broadcast
  channels,'' in \emph{Proc. {IEEE} Int. Symp. Personal Indoor Mobile Radio
  Commun. (PIMRC)}, Oct. 2020, pp. 1--6.

\bibitem{Zhang2019CRS}
J.~Zhang, B.~Clerckx, J.~Ge, and Y.~Mao, ``{Cooperative rate splitting for MISO
  broadcast channel with user relaying, and performance benefits over
  cooperative NOMA},'' \emph{{IEEE} Signal Process. Lett.}, vol.~26, no.~11,
  pp. 1678--1682, Sept. 2019.

\bibitem{lu2021secureSWIPT}
Y.~Lu, K.~Xiong, P.~Fan, Z.~Zhong, B.~Ai, and K.~B. Letaief, ``Worst-case
  energy efficiency in secure {SWIPT} networks with rate-splitting {ID} and
  power-splitting {EH} receivers,'' \emph{{IEEE} Trans. Wireless Commun.},
  vol.~21, no.~3, pp. 1870--1885, Mar. 2022.

\bibitem{cai2021secureRA}
T.~Cai, J.~Zhang, S.~Yan, L.~Meng, J.~Sun, and N.~Al-Dhahir, ``Resource
  allocation for secure rate-splitting multiple access with adaptive
  beamforming,'' in \emph{Proc. {IEEE} Int. Conf. Commun. (ICC) Workshop}, Jul.
  2021, pp. 1--6.

\bibitem{bruno2020ZF}
B.~Clerckx, Y.~Mao, R.~Schober, and H.~V. Poor, ``{Rate-splitting unifying
  SDMA, OMA, NOMA, and multicasting in MISO broadcast channel: A simple
  two-user rate analysis},'' \emph{{IEEE} Wireless Commun. Lett.}, vol.~9,
  no.~3, pp. 349--353, Mar. 2020.

\bibitem{mukherjee2014PLSsurvey}
A.~{Mukherjee}, S.~A.~A. {Fakoorian}, J.~{Huang}, and A.~L. {Swindlehurst},
  ``Principles of physical layer security in multiuser wireless networks: A
  survey,'' \emph{{IEEE} Commun. Surveys Tuts.}, vol.~16, no.~3, pp.
  1550--1573, 3rd Quart, 2014.

\bibitem{LYC1978Gaussian}
S.~Leung-Yan-Cheong and M.~Hellman, ``{The Gaussian wire-tap channel},''
  \emph{{IEEE} Trans. Inf. Theory}, vol.~24, no.~4, pp. 451--456, Jul. 1978.

\bibitem{boyd2004convex}
S.~Boyd and L.~Vandenberghe, \emph{Convex Optimization}.\hskip 1em plus 0.5em
  minus 0.4em\relax Cambridge, U.K.: Cambridge Univ. Press, 2004.

\bibitem{Hamdi2016RSWMMSE}
H.~Joudeh and B.~Clerckx, ``Sum-rate maximization for linearly precoded
  downlink multiuser {MISO} systems with partial {CSIT}: A rate-splitting
  approach,'' \emph{{IEEE} Trans. Commun.}, vol.~64, no.~11, pp. 4847--4861,
  Nov. 2016.

\bibitem{razaviyayn2014successive}
M.~Razaviyayn, ``Successive convex approximation: Analysis and applications,''
  Ph.D. dissertation, University of Minnesota, 2014.

\bibitem{razaviyayn2013unified}
M.~Razaviyayn, M.~Hong, and Z.-Q. Luo, ``A unified convergence analysis of
  block successive minimization methods for nonsmooth optimization,''
  \emph{SIAM Journal on Optimization}, vol.~23, no.~2, pp. 1126--1153, 2013.

\bibitem{Christensen2008WMMSE}
S.~S. Christensen, R.~Agarwal, E.~De~Carvalho, and J.~M. Cioffi, ``{Weighted
  sum-rate maximization using weighted MMSE for MIMO-BC beamforming design},''
  \emph{{IEEE} Trans. Wireless Commun.}, vol.~7, no.~12, pp. 4792--4799, Dec.
  2008.

\bibitem{Shi2011WeightedMMSE}
Q.~Shi, M.~Razaviyayn, Z.-Q. Luo, and C.~He, ``{An iteratively weighted MMSE
  approach to distributed sum-utility maximization for a MIMO interfering
  broadcast channel},'' \emph{{IEEE} Trans. Signal Process.}, vol.~59, no.~9,
  pp. 4331--4340, Sept. 2011.

\end{thebibliography}

\balance
\end{document}